\documentclass[journal,twoside,web]{ieeecolor}
\usepackage{tmi}
\usepackage{cite}
\usepackage{amsmath,amssymb,amsfonts}
\usepackage{algorithmicx}
\usepackage{graphicx}
\usepackage{textcomp}

\usepackage{algorithm}
\usepackage{xspace}
\usepackage{booktabs}
\usepackage{multirow}
\usepackage{makecell}
\usepackage{mathrsfs}
\usepackage{algpseudocode}
\usepackage{makecell}
\usepackage{amsmath}

\usepackage{hyperref}
\hypersetup{hypertex=true,
            colorlinks=true,
            linkcolor=blue,
            anchorcolor=blue,
            citecolor=blue}

\newcommand{\methodname}{CoreDiff\xspace}
\newcommand{\methodvariantp}{CoreDiff+OSL\textsuperscript{p}\xspace}
\newcommand{\methodvariantu}{CoreDiff+OSL\textsuperscript{u}\xspace}
\newcommand{\networkname}{CLEAR-Net\xspace}

\newcommand{\vct}[1]{\boldsymbol{#1}} 
\newcommand{\fun}[1]{\mathcal{#1}} 
\newcommand{\etal}{\textit{et al}.\xspace}
\newcommand{\ie}{\textit{i}.\textit{e}.\xspace}
\newcommand{\eg}{\textit{e}.\textit{g}.\xspace}

\usepackage{tikz}
\usepackage{textcomp}
\usepackage{lipsum}
\newcommand\copyrighttext{%
  \footnotesize \textcopyright 2023 IEEE. Personal use of this material is permitted.
  Permission from IEEE must be obtained for all other uses, in any current or future
  media, including reprinting/republishing this material for advertising or promotional
  purposes, creating new collective works, for resale or redistribution to servers or
  lists, or reuse of any copyrighted component of this work in other works.
  DOI: \href{https://doi.org/10.1109/TMI.2023.3320812}{10.1109/TMI.2023.3320812}.}
\newcommand\copyrightnotice{%
\begin{tikzpicture}[remember picture,overlay]
\node[anchor=south,yshift=3pt] at (current page.south) {\fbox{\parbox{\dimexpr\textwidth-\fboxsep-\fboxrule\relax}{\copyrighttext}}};
\end{tikzpicture}%
}

\usepackage{accents}
\newlength{\dhatheight}
\newcommand{\doublehat}[1]{%
    \settoheight{\dhatheight}{\ensuremath{\widehat{#1}}}%
    \addtolength{\dhatheight}{-0.25ex}%
    \widehat{\vphantom{\rule{1pt}{\dhatheight}}%
    \smash{\widehat{#1}}}}

\newlength\savewidth\newcommand\shline{\noalign{\global\savewidth\arrayrulewidth
  \global\arrayrulewidth 1.25pt}\hline\noalign{\global\arrayrulewidth\savewidth}}

\def\BibTeX{{\rm B\kern-.05em{\sc i\kern-.025em b}\kern-.08em
    T\kern-.1667em\lower.7ex\hbox{E}\kern-.125emX}}
\begin{document} 
\title{\methodname: Contextual Error-Modulated Generalized Diffusion Model for Low-Dose CT Denoising and Generalization}

\author{
Qi~Gao,~\IEEEmembership{Graduate Student Member, IEEE},
Zilong~Li,~\IEEEmembership{Graduate Student Member, IEEE},
Junping~Zhang,~\IEEEmembership{Senior Member, IEEE},
Yi~Zhang,~\IEEEmembership{Senior Member, IEEE} and
Hongming~Shan,~\IEEEmembership{Senior Member, IEEE}
\thanks{Q. Gao and H. Shan are with the Institute of Science and Technology for Brain-inspired Intelligence and MOE Frontiers Center for Brain Science, Fudan University, Shanghai 200433, China, and also with Shanghai Center for Brain Science and Brain-inspired Technology, Shanghai 201602, China (e-mail: qgao21@m.fudan.edu.cn; hmshan@fudan.edu.cn)}
\thanks{Z. Li and J. Zhang are with the Shanghai Key Lab of Intelligent Information Processing, School of Computer Science, Fudan University, Shanghai 200433, China (e-mail: zilongli21@m.fudan.edu.cn; jpzhang@fudan.edu.cn)}
\thanks{Y. Zhang is with School of Cyber Science and Engineering, Sichuan University, Chengdu, Sichuan 610065, China (e-mail: yzhang@scu.edu.cn)}
}

\maketitle
\copyrightnotice

\begin{abstract}
Low-dose computed tomography (CT) images suffer from noise and artifacts due to photon starvation and electronic noise.
Recently, some works have attempted to use diffusion models to address the over-smoothness and training instability encountered by previous deep-learning-based denoising models.
However, diffusion models suffer from long inference time due to a large number of sampling steps involved.
Very recently, cold diffusion model generalizes classical diffusion models and has greater flexibility. 
Inspired by cold diffusion, this paper presents a novel COntextual eRror-modulated gEneralized Diffusion model for low-dose CT (LDCT) denoising, termed \methodname. 
First, \methodname utilizes LDCT images to displace the random Gaussian noise and employs a novel mean-preserving degradation operator to mimic the physical process of CT degradation, significantly reducing sampling steps thanks to the informative LDCT images as the starting point of the sampling process.
Second, to alleviate the error accumulation problem caused by the imperfect restoration operator in the sampling process, we propose a novel ContextuaL Error-modulAted Restoration Network (\networkname), which can leverage contextual information to constrain the sampling process from structural distortion and modulate time step embedding features for better alignment with the input at the next time step.
Third, to rapidly generalize the trained model to a new, unseen dose level with as few resources as possible, we devise a one-shot learning framework to make \methodname generalize faster and better using only one single LDCT image (un)paired with normal-dose CT (NDCT).
Extensive experimental results on four datasets demonstrate that our \methodname outperforms competing methods in denoising and generalization performance, with clinically acceptable inference time. Source code is made available at \url{https://github.com/qgao21/CoreDiff}.
\end{abstract}

\begin{IEEEkeywords}
Low-dose CT, denoising, diffusion model, one-shot learning.
\end{IEEEkeywords}
\section{Introduction}
\label{sec:introduction}
\IEEEPARstart{C}{omputed} tomography (CT) is a widely-used imaging modality in clinical diagnosis. However, X-ray ionizing radiation in CT scans could cause health risks such as hair loss and cancer~\cite{smith2009radiation, sodickson2009recurrent}. 
One can reduce the radiation dose by lowering the tube current in clinical practice. 
Unfortunately, the resulting low-dose CT (LDCT) images contain severe noise and artifacts, compromising the radiologists' diagnosis. 
When the raw data are accessible, vendor-specific sinogram preprocessing or iterative reconstruction algorithms can effectively remove noise from LDCT images. However, sinogram preprocessing may cause blurred edges and resolution loss while iterative reconstruction methods suffer from expensive computational cost~\cite{xie2017robust, wang2006penalized, shan2018d}. In addition, raw data are typically not available to researchers due to commercial privacy. 
Unlike them, image post-processing algorithms~\cite{aharon2006k, feruglio2010block, chen2013improving} directly process the reconstructed images and are gaining popularity due to their plug-and-play nature without access to raw data. For example, Ma~\etal leveraged the redundancy of information in the previous normal-dose scan to compute non-local weights for non-local mean (NLM)-based LDCT image denoising~\cite{ma2011low}. Li~\etal utilized the analytical noise map obtained from repeated scans of the phantom data to improve the NLM algorithm, enabling adaptive denoising based on the local noise level of the CT image~\cite{li2014adaptive}. Sheng~\etal proposed a block matching 3D (BM3D)-based algorithm for low-dose megavoltage CT image denoising, which uses a saliency map, derived from the residual texture information after BM3D denoising, to enhance the visual conspicuity of soft tissue~\cite{sheng2014denoised}.

In recent years, many efforts have been made to develop deep learning (DL) techniques for LDCT image post-processing, achieving promising performance~\cite{shan2019competitive, wang2020deep}. 
Initially, some researchers optimized encoder-decoder networks by minimizing the pixel-wise loss between the denoised and normal-dose CT (NDCT) images; one representative model is the residual encoder-decoder convolutional neural network (RED-CNN)~\cite{chen2017low}. 
Further, Xia~\etal integrated RED-CNN into a parameter-dependent framework (PDF-RED-CNN) for multiple geometries and dose levels~\cite{xia2021ct}. Despite the outstanding denoising performance, these methods often lead to over-smoothing images~\cite{kim2019performance, nagare2021bias}. To alleviate this problem, some works use generative adversarial networks (GANs) to preserve more textures and details as close to NDCT images as possible~\cite{kang2019cycle}. For example, Yang~\etal combined Wasserstein GAN  and perceptual loss (WGAN-VGG) to produce more realistic denoised images~\cite{yang2018low}. Huang~\etal proposed a dual-domain GAN (DU-GAN) to learn the global and local differences between the denoised and NDCT images~\cite{huang2021gan}. However, GANs are usually difficult to train due to their adversarial nature, and require careful design of optimization and network architectures to ensure convergence~\cite{zhao2018bias}. 

Recently, diffusion models have received much attention due to their impressive image generation performance~\cite{sohl2015deep, ho2020denoising, song2020denoising, nichol2021improved, ye_3dinverse, ye_improving}, enjoying the advantages of multiple generative models: good distribution coverage similar to variational autoencoder and better generation quality than GANs~\cite{song2020score, dhariwal2021diffusion, yang2022diffusion}. 
However, since the diffusion models generate images progressively from Gaussian noise, they suffer from expensive computational cost for inference due to the multiple iterative sampling; \eg denoising diffusion probabilistic model (DDPM)~\cite{ho2020denoising} requires 1,000 sampling steps. 
This limits their application in various real-time scenarios, especially in the field of medical imaging~\cite{gao2022cocodiff, xia2022low}. 
Some works accelerate diffusion models to make them practical. For example, Nichol and Dhariwal enhanced the log-likelihood performance of the DDPM and reduced the sampling steps to 100~\cite{nichol2021improved}. 
Xia~\etal used a fast ordinary differential equation solver to accelerate the DDPM for LDCT image denoising, requiring only 50 sampling steps~\cite{xia2022low}. 
Despite inference time being reduced to some extent, these improved diffusion models focus on the trade-off between performance and sampling speed within the theoretical framework of classical diffusion models. 
Recently, a generalized diffusion model, referred to as cold diffusion, extends classical diffusion models by gradually degrading images through a pre-defined degradation operator, such as adding various types of noise, blurring, downsampling, etc~\cite{bansal2022cold, yen2022cold}. Cold diffusion uses a learnable restoration operator to reverse the diffusion process and generates images through a ``restoration-redegradation'' sampling process. 
Although cold diffusion allows customizing the diffusion process, its performance is subject to the learned restoration operator. 
In practice, the learned restoration operator may be imperfect, leading to accumulated errors between the restored and ground truth images after multiple sampling iterations and causing non-negligible pixel-wise deviations.

In this paper, we propose a contextual error-modulated generalized diffusion model (\methodname) for LDCT denoising inspired by cold diffusion. 
To accelerate sampling, we develop a mean-preserving degradation operator applicable to the LDCT denoising task with the LDCT images as the endpoint of the diffusion process (forward) and the starting point of the sampling process (reverse). In doing so, the number of sampling steps can be significantly reduced since LDCT images (warm state) are more informative than random Gaussian noise (hot state). 
To alleviate the accumulated error caused by imperfect restoration operators during the sampling process, we further propose a novel contextual error-modulated restoration network (\networkname), which can 
leverage rich contextual information from adjacent slices to mitigate structural distortion in z-axis, and rectify the misalignment between the input image and time-step embedding features through an error-modulated module.
Finally, benefiting from the proposed mean-preserving degradation operator, we devise a one-shot learning framework, which can quickly generalize \methodname to a new, unseen dose level using one single LDCT image (un)paired with NDCT.  

In summary, the contributions of this work are listed as follows.
\emph{First}, we propose a novel generalized diffusion model \methodname for LDCT denoising, in which the resulting diffusion process mimics the physical process of CT image degradation.
To the best of our knowledge, this is the ﬁrst work to extend the cold diffusion model for LDCT denoising.
\emph{Second}, we introduce a novel restoration network \networkname, which can mitigate accumulated errors by constraining the sampling process using contextual information among adjacent slices and calibrating the time step embedding feature using the latest prediction.
\emph{Third}, we further devise a one-shot learning framework, which can quickly and easily adapt the trained \methodname to a new, unseen dose level. This can be done with one single LDCT image (un)paired with NDCT.  
\emph{Fourth}, extensive experiment results on four test datasets demonstrate the superior performance of the proposed \methodname, with a clinically acceptable inference time of 0.12 seconds per slice.
\begin{figure*}[!t]
\centerline{\includegraphics[width=1\linewidth]{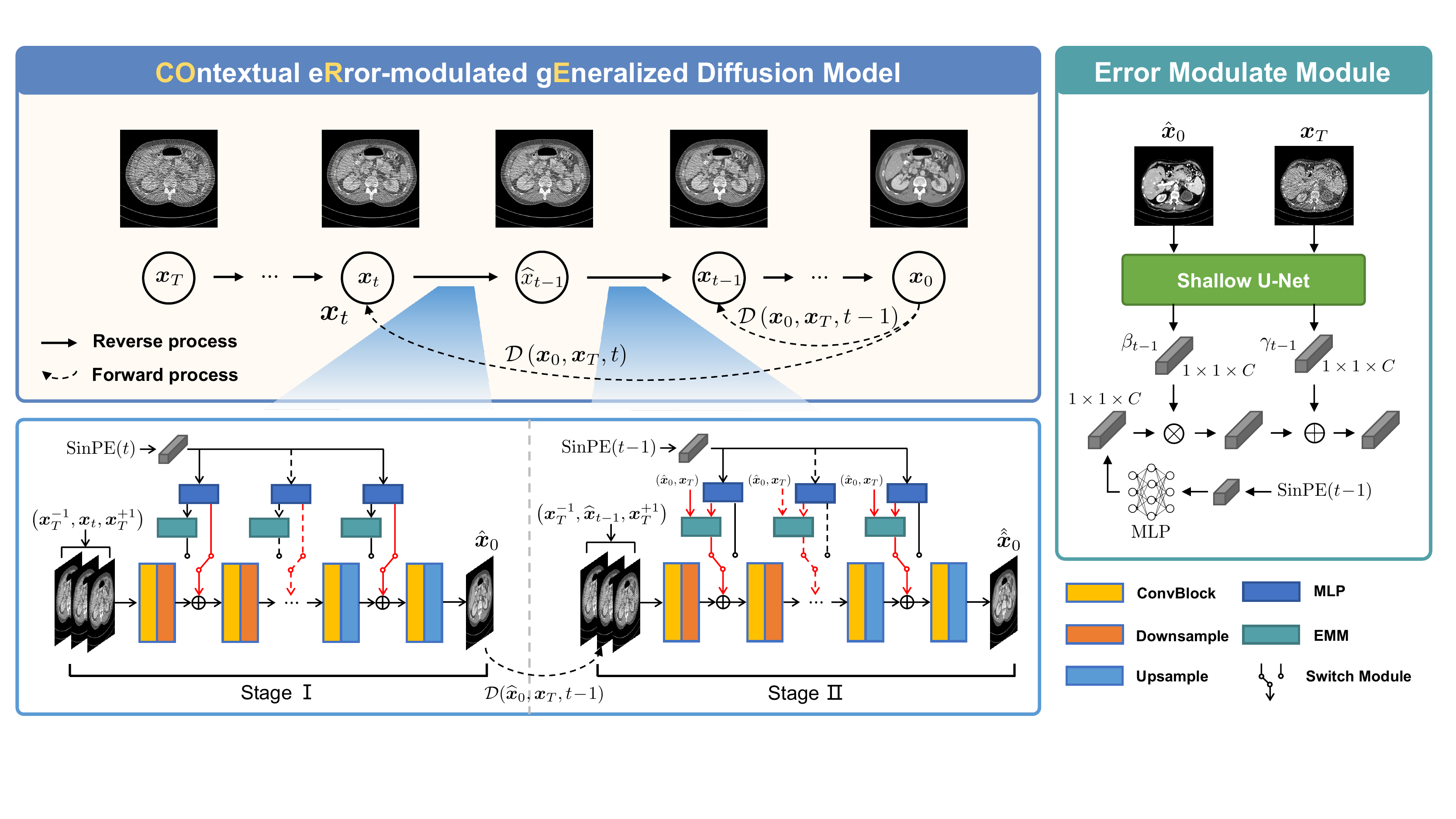}}
\caption{Overview of the proposed \methodname for low-dose CT denoising. The introduced generalized diffusion model leverages a novel degradation operator to mimic the physical process of CT image degradation during the diffusion process. The proposed \networkname can alleviate the accumulated error and is trained in a two-stage manner for each time step; one key feature of \networkname is the error-modulated module (EMM) that can calibrate the time step embedding feature with the latest prediction and the given input LDCT image.}
\label{fig:overall_framwork}
\end{figure*}

\begin{figure}[!t]\centerline{
\includegraphics[width=1.0\linewidth]{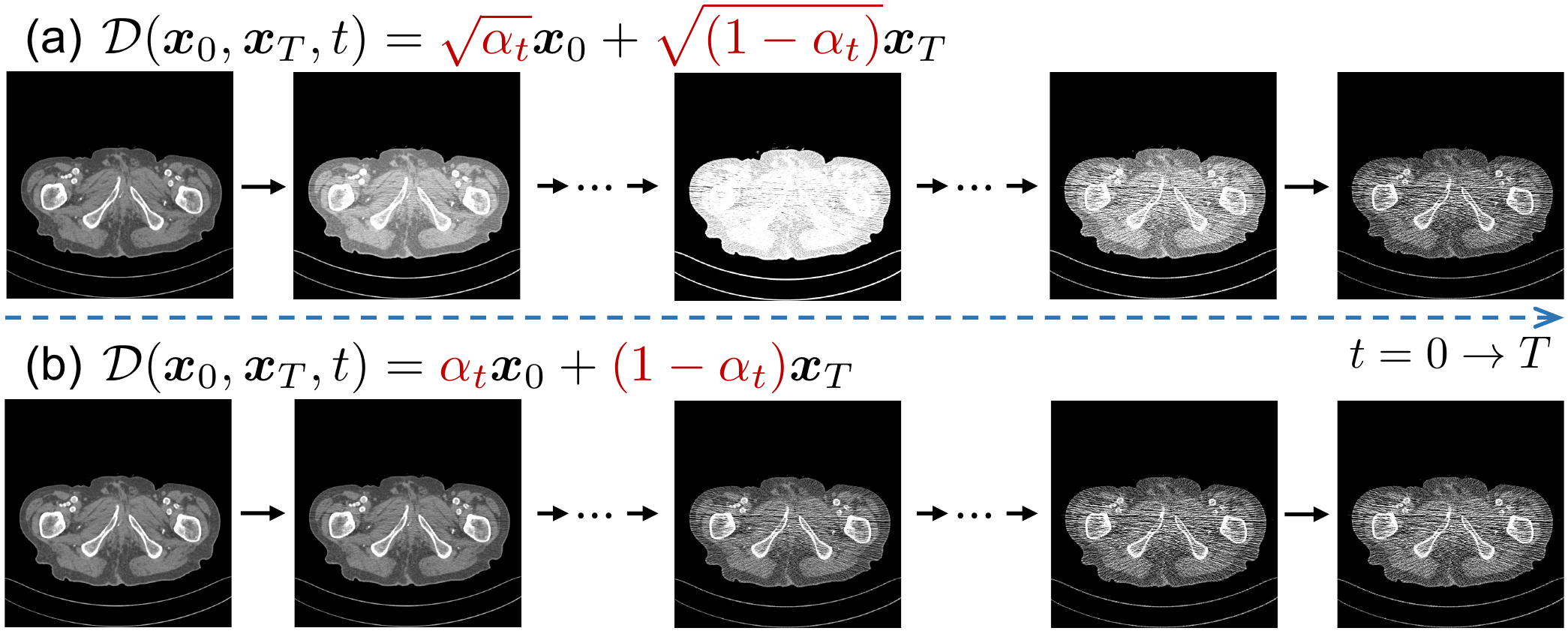}}
\caption{Comparison of (a) the degradation operator in Eq.~\eqref{eq:commonly_used_degradation_operator}  and (b) the proposed one in Eq.~\eqref{eq:proposed_degradation_operator}. The proposed operator achieves a mean-preserving process, simulating the physical process of CT degradation. }
\label{fig:different_diffusion_process}
\end{figure}

\section{Method}
In this section, we first introduce the basic principles of the cold diffusion model and the error accumulation issue. Then we present our \methodname for LDCT denoising in a generalized diffusion model framework with a new mean-preserving diffusion process and a new contextual error-modulated restoration network (CLEAR-Net), followed by a one-shot learning framework for rapid generalization.

\subsection{Preliminaries: Cold Diffusion}\label{sec:Cold_Diffusion}
Cold diffusion model is a generalized diffusion model~\cite{bansal2022cold},
which extends the diffusion and sampling of Gaussian noise to any type of degradation such as adding various types of noise, blurring, downsampling, etc.
Specifically, given an image $\vct{x}_{0}$ from the training data distribution $Q$, a customized \emph{degradation} operator $\fun{D}(\cdot)$ is used to gradually degrade the image $\vct{x}_{0}$ (cold state) into the image $\vct{x}_{T}$ sampled from a random initial distribution $P$ (hot state), \eg Gaussian distribution, where $T$ is the total number of time steps for diffusion.  Then, image $\vct{x}_{t}$ of any time step $t$ during the diffusion process is defined as $
\vct{x}_{t} = \fun{D}(\vct{x}_{0}, \vct{x}_{T}, t)$, where $t$ corresponds to the degree of degradation and the operator $\fun{D}(\cdot)$ should be continuous for any $t$. In the context of LDCT denoising, adding noise is the most related degradation operator. For the diffusion process of adding noise, the degradation operator in the cold diffusion model is the same as the one used in classical diffusion models, defined as:
\begin{align}
\vct{x}_{t}=\fun{D}(\vct{x}_{0}, \vct{x}_{T}, t)=\sqrt{\alpha_{t}} \vct{x}_{0}+\sqrt{(1-\alpha_{t})} \vct{x}_{T},
\label{eq:commonly_used_degradation_operator}
\end{align}
where  $\vct{x}_{T}$ is random noise with known distribution and $\alpha_t<\alpha_{t-1}$, $\forall\  1\leq t \leq T$.

In the reverse process, we first sample $\vct{x}_{T}$ from $P$, and then use a \emph{restoration} operator $\fun{R}(\cdot)$ to reverse the diffusion process, which can be expressed as follows:
\begin{align}
\widehat{\vct{x}}_{0} = \fun{R}(\vct{x}_{t}, t) \approx \vct{x}_{0}.
\label{eq:cold_diff_restoration_operator}
\end{align}
In practice, $\fun{R}(\cdot)$ is a neural network parameterized by  $\vct{\theta}$, which can be optimized by the following objective function:
\begin{align}
\min_{\vct{\theta}} \mathbb{E}_{\vct{x}_{0} \sim Q, \vct{x}_{T} \sim P}\|\fun{R}_{\vct{\theta}}(\fun{D}(\vct{x}_{0}, \vct{x}_{T}, t), t)-\vct{x}_{0}\|.
\label{eq:cold_diff_optimization}
\end{align}
Note that for any $t$, $\fun{R}_{\vct{\theta}}(\cdot)$ can directly generate the restored image $\widehat{\vct{x}}_{0}$ from $\vct{x}_{T}$. However, we highlight that such a one-step prediction could produce blurred image $\widehat{\vct{x}}_{0}$ with severe detail loss~\cite{bansal2022cold}.

To address this, following the annealing sampling algorithm in classical diffusion models~\cite{ho2020denoising, song2020score, nichol2021improved}, the cold diffusion model uses a ``\emph{restoration}-\emph{redegradation}'' sampling algorithm to gradually generate images with a total of $T$ sampling steps. Image $\widehat{\vct{x}}_{t-1}$ at time step $t-1$ can be calculated based on the prediction $\widehat{\vct{x}}_{0}$ as follows:
\begin{align}
\widehat{\vct{x}}_{t-1}=\fun{D}(\widehat{\vct{x}}_{0}, \vct{x}_{T}, t\!-\!1).
\label{eq:redegradation}
\end{align}
Although such an iterative sampling algorithm can produce sharper images than a one-step prediction, the prediction error between $\vct{x}_{0}$ and $\widehat{\vct{x}}_{0}$ could introduce misalignment between $\widehat{\vct{x}}_{t-1}$ and time step $t-1$. As a result, the prediction bias of $\fun{R}_{\vct{\theta}}(\cdot)$ may further worsen by the misalignment, as errors are accumulated during the sampling process.
Bansal~\etal~proposed an improved sampling algorithm to reduce this accumulated error~\cite{bansal2022cold}:
\begin{align}
\vct{x}_{t-1} = \vct{x}_{t}-\fun{D}(\widehat{\vct{x}}_{0}, \widehat{\vct{x}}_{T}, t)+\fun{D}(\widehat{\vct{x}}_{0}, \widehat{\vct{x}}_{T}, t\!-\!1),
\label{eq:cold_diff_improve_sample}
\end{align}
where $\widehat{\vct{x}}_{T}=\left(\vct{x}_{t}-\sqrt{\alpha_{t}} \widehat{\vct{x}}_{0}\right) / \sqrt{\left(1-\alpha_{t}\right)}$. Although the improved sampling algorithm in Eq.~\eqref{eq:cold_diff_improve_sample} mitigates the issue of error accumulation and has been shown to produce better image quality~\cite{bansal2022cold, yen2022cold},
it does not rectify the misalignment between the input and its corresponding time step, which could cause a non-negligible shift in the pixel value.

\subsection{The Proposed \methodname Model}\label{sec:Proposed Model}
Fig.~\ref{fig:overall_framwork} presents the overall architecture of our \methodname, which involves a generalized diffusion model with LDCT images as the endpoint of the diffusion process, a new mean-preserving degradation operator to mimic the physical process of CT degradation, and a novel \networkname to address the accumulated errors and misalignment in cold diffusion.

\begin{algorithm}[!t]
\caption{Training for \methodname}
\label{alg_training}
\begin{algorithmic}[1]
\Require Paired ND/LDCT image sets ${I=\{\left(\vct{x}_{0}, \vct{x}_{T}\right)_{i}\}_{i=1}^{N}}$, total time steps $T$
\Ensure Trained \networkname $\fun{R}_{\vct{\theta}}$
\State \textbf{Initialization:} Randomly initializes \networkname $\fun{R}_{\vct{\theta}}$
\Repeat
\State \textbf{Sample} ${(\vct{x}_{0}, \vct{x}_{T}) \sim I}$
\State \textbf{Sample} ${t \sim \operatorname{Uniform}(\{1, \ldots, T\})}$
\State \textbf{Calculate} $\vct{x}_{t}^{\text{c}}$ by Eq.~\eqref{eq:proposed_degradation_operator} and $\operatorname{Concat}(\cdot)$\Comment{Stage \uppercase\expandafter{\romannumeral1}}
\State ${\widehat{\vct{x}}_{0} \leftarrow \fun{R}_{\vct{\theta}}\left(\vct{x}_{t}^{\text{c}}, t\right)}$
\State \textbf{Calculate} $\widehat{\vct{x}}_{t-1}^{\text{c}}$ by Eq.~\eqref{eq:redegradation} and $\operatorname{Concat}(\cdot)$\Comment{Stage \uppercase\expandafter{\romannumeral2}}
\State $\doublehat{\vct{x}}_{0} \leftarrow \fun{R}_{\vct{\theta}}\left(\widehat{\vct{x}}_{t-1}^{\text{c}}, t\!-\!1, \fun{F}_{\vct{\phi}}\left(\widehat{\vct{x}}_{0}, \vct{x}_{T}\right)\right)$
\State \textbf{Update} $\vct{\theta},\vct{\phi}$ by Eq.~\eqref{eq:loss_function}
\Until{converged}
\end{algorithmic}
\end{algorithm}

\subsubsection{Generalized diffusion model for low-dose CT}
Previous diffusion-based LDCT denoising methods~\cite{gao2022cocodiff, xia2022low}  typically characterize the diffusion process as the addition of Gaussian noise and use LDCT images as a condition to predict the corresponding NDCT image. However, it is important to note that the statistical characteristics of noise in CT images are complex and cannot be simply modelled by a Gaussian distribution. 
Moreover, for noise with zero means, it is commonly assumed that a clean image represents the expectation of multiple sets of noise measurements~\cite{joshi2010seeing, lehtinen2018noise2noise, wu2019consensus}. In the context of the LDCT denoising task, we consider that the NDCT image $\vct{x}_{0}$ represents the expectation of a collection of its LDCT counterparts $\{\vct{x}_{T}^{i}\}_{i=1}^N$. However, as shown in Fig.~\ref{fig:different_diffusion_process}(a), we find that because the sum of $\sqrt{\alpha_{t}}$ and $\sqrt{(1-\alpha_{t})}$ is not consistently equal to 1, the expectation of intermediate images calculated by Eq.~\eqref{eq:commonly_used_degradation_operator} deviates from $\vct{x}_{0}$ and exhibits obvious CT number drifts during the diffusion processes. Therefore, the widely-adopted degradation operator in Eq.~\eqref{eq:commonly_used_degradation_operator} departs from the actual physical process of CT degradation due to the dose reduction.

Unlike previous diffusion-based methods that transform the LDCT denoising task into a conditional image generation task with random Gaussian noise as the end point of the diffusion process and require a large number of steps to generate an accurately estimated image, we propose a generalized diffusion model for LDCT denoising, which uses LDCT images as the endpoint of the diffusion process, \ie $\vct{x}_{T}$. To make the diffusion process mimic the physical process of CT image degradation, we introduce a new degradation operator $\fun{D}(\cdot)$ defined as follows:
\begin{align}
\vct{x}_{t} = \fun{D}(\vct{x}_{0}, \vct{x}_{T}, t)={\alpha_{t}} \vct{x}_{0}+{(1-\alpha_{t})} \vct{x}_{T},
\label{eq:proposed_degradation_operator}
\end{align}
where image $\vct{x}_{t}$ at each time step retains the noise statistics specific to LDCT image $\vct{x}_{T}$. 
As shown in Fig.~\ref{fig:different_diffusion_process}(b), another merit of employing this operator is its capability to ensure that the intermediate image $\vct{x}_{t}$ of the diffusion process maintains the same expectation $\vct{x}_{0}$, without introducing additional CT number shifts. 
Therefore, we refer to this degradation operator as the mean-preserving degradation operator; we note that in a practical scenario, it may not be strictly mean-preserving due to complicated noise and the presence of artifacts. 
Such property not only makes the diffusion process of \methodname consistent with the LDCT image degradation process but also is important for our one-shot learning framework. The LDCT image $\vct{x}_{T}$ can be considered as an intermediate state between the cold state (clean image) and the hot state (random noise), which we refer to as the warm state. 
As described in Sec.~\ref{sec:Cold_Diffusion}, the sampling process of cold diffusion and classical Gaussian diffusion models starts from random Gaussian noise and progressively diminishes the noise of the image until the $\widehat{\vct{x}}_{0}$ is generated. Therefore, they require a large number of sampling steps to generate an image with a noise level similar to that of a LDCT image, which contains the fundamental semantic information of the NDCT image. As a result, the proposed \methodname can perform sampling from the warm state using a smaller $T$, instead of starting from random Gaussian noise.

\begin{algorithm}[!t]
\caption{Sampling for \methodname}
\label{alg_sampling}
\begin{algorithmic}[1]
\Require A test LDCT image $\vct{x}_{T}$
\Ensure Denoised image $\vct{x}_{0}$
\State \textbf{Load} the trained \networkname $\fun{R}_{\vct{\theta}}$
\State $\vct{x}_{T}^{\text{c}} \leftarrow \operatorname{Concat}(\vct{x}_{T}^{-1},\vct{x}_{T}, \vct{x}_{T}^{+1})$
\For{${t=T, T-1, \ldots, 1}$}
\State $\vct{x}_{0} \leftarrow \fun{R}_{\vct{\theta}}\left(\vct{x}_{t}^{c}, t , \fun{F}_{\vct{\phi}}\left(\vct{x}_{0}, \vct{x}_{T}\right)\right)$
\State \textbf{Calculate} $\vct{x}_{t-1}^{c}$ by Eq.~\eqref{eq:cold_diff_improve_sample} and $\operatorname{Concat}(\cdot)$
\EndFor
\end{algorithmic}
\end{algorithm}

\subsubsection{Contextual Error-modulated Restoration Network (\networkname)}
To mitigate accumulated errors and the misalignment in cold diffusion caused by an imperfect restoration network, we introduce a novel restoration network, called \networkname. Based on the ``restoration-redegradation'' sampling algorithm, we split each time step in the training process into two stages, as shown in Fig.~\ref{fig:overall_framwork}: 1) in Stage \uppercase\expandafter{\romannumeral1}, we first obtain the degraded image $\vct{x}_{t}$ using Eq.~\eqref{eq:proposed_degradation_operator}, and then use \networkname, $\fun{R}_{\vct{\theta}}(\cdot)$, to estimate $\widehat{\vct{x}}_{0}$; and 2)
in Stage \uppercase\expandafter{\romannumeral2}, we perform the redegradation operation in Eq.~\eqref{eq:redegradation} to compute $\widehat{\vct{x}}_{t-1}$ based on the latest prediction $\widehat{\vct{x}}_{0}$, and then use the same network, $\fun{R}_{\vct{\theta}}(\cdot)$,  to predict the NDCT image.
The novelties of our \networkname are two-fold. On the one hand, inspired by the contextual information used in~\cite{gao2022cocodiff, shan2018d}, we introduce the contextual information from adjacent slices of $\vct{x}_{T}$ to mitigate the structural distortion during the sampling process.
More specifically, we assume the adjacent slices of $\vct{x}_{T}$ are $\vct{x}_{T}^{-1}$ and $\vct{x}_{T}^{+1}$, corresponding to its upper and lower slice, respectively. We concatenate $\vct{x}_{t} \in \mathbb{R}^{1\times H\times W}$ at each step and the adjacent slices at the starting point, $\vct{x}_{T}^{-1}$ and $\vct{x}_{T}^{+1}$, along the channel dimension, which yields a contextual version of $\vct{x}_{t}$, \ie $\vct{x}_{t}^{\text{c}}=\operatorname{Concat}(\vct{x}_{T}^{-1},\vct{x}_{t}, \vct{x}_{T}^{+1})\in \mathbb{R}^{3\times H\times W}$. Since the adjacent slices remain unchanged during sampling, thereby constraining the network $\fun{R}_{\vct{\theta}}(\cdot)$ to produce continuous z-axis structures.

On the other hand, \networkname leverages an error-modulated module (EMM) to calibrate the misalignment between the input to the network ${\widehat{\vct{x}}_{t-1}}$  and the time-step embedding features of $t-1$. 
Specifically, our EMM is a feature-wise linear modulation module~\cite{perez2018film, huang2017arbitrary, dumoulin2016learned} to modulate the time step embedding features, in which the modulation factors at time step ${t-1}$ are estimated as follows:
\begin{align}
\vct{\beta}_{t-1}, \vct{\gamma}_{t-1}=\fun{F}_{\vct{\phi}}(\widehat{\vct{x}}_{0}, \vct{x}_{T}),\quad \widehat{\vct{x}}_{0}=\fun{R}_{\vct{\theta}}(\vct{x}_{t}, t),
\label{eq:modulation_factor_calculate}
\end{align}
where $\fun{F}_{\vct{\phi}}(\cdot)$ is a shallow network parameterized by $\vct{\phi}$ to estimate the modulation factors based on the latest prediction $\widehat{\vct{x}}_{0}$ and the initial input LDCT image $\vct{x}_{T}$.
Then the time step embedding features of $t-1$ are modulated as follows:
\begin{align}
\widetilde{\vct{f}}_{t-1}\!=\!\vct{\beta}_{t-1} \vct{f}_{t-1}+\vct{\gamma}_{t-1},\ \vct{f}_{t-1}\!=\!\operatorname{MLP}(\operatorname{SinPE}(t\!-\!1)),
\label{eq:modulation_feature}
\end{align}
where $\operatorname{SinPE}(\cdot)$ represents Sinusoidal position encoding for time step $t-1$, $\vct{f}_{t-1}$ is the time step embedding feature from a multi-layer perceptron (MLP), and $\widetilde{\vct{f}}_{t-1}$ is the modulated one. Note that the proposed EMM is only involved in Stage \uppercase\expandafter{\romannumeral2} and the modulated features are used after each up-/down-sampling operation in $\fun{R}_{\vct{\theta}}(\cdot)$.

With the proposed \networkname,  the final training objective of our \methodname is defined as follows:
\begin{align}
\min_{\vct{\theta}, \vct{\phi}}\ \mathbb{E}\Big[\|\underbrace{\fun{R}_{\vct{\theta}}(\vct{x}_{t}^{\text{c}}, t)-\vct{x}_{0}}_\text{Stage~\uppercase\expandafter{\romannumeral1}}\|^{2} +\|\underbrace{\doublehat{\vct{x}}_{0}-\vct{x}_{0}}_\text{Stage~\uppercase\expandafter{\romannumeral2}}\|^{2}\Big],
\label{eq:loss_function}
\end{align}
where $\doublehat{\vct{x}}_{0}$ denotes the output of $\fun{R}_{\vct{\theta}}(\cdot)$ at time step $t-1$ in Stage \uppercase\expandafter{\romannumeral2}; \ie $\doublehat{\vct{x}}_{0} = \fun{R}_{\vct{\theta}}(\widehat{\vct{x}}_{t-1}^{\text{c}}, t\!-\!1, \fun{F}_{\vct{\phi}}(\widehat{\vct{x}}_{0}, \vct{x}_{T}))$ and $\widehat{\vct{x}}_{t-1}^{\text{c}}=\operatorname{Concat}(\vct{x}_{T}^{-1},\widehat{\vct{x}}_{t-1}, \vct{x}_{T}^{+1})$.

Finally, in the sampling process of \methodname, the degradation operator and the restoration operator are performed only once at each time step. With the trained \networkname, we use the improved sampling algorithm in Eq.~\eqref{eq:cold_diff_improve_sample} and replace the coefficients according to our degradation operator in Eq.~\eqref{eq:proposed_degradation_operator} to infer the final denoised image. The training and sampling (inference) procedures are shown in Algs.~\ref{alg_training} and~\ref{alg_sampling}, respectively.

\begin{figure}[!t]
\centerline{
\includegraphics[width=1\linewidth]{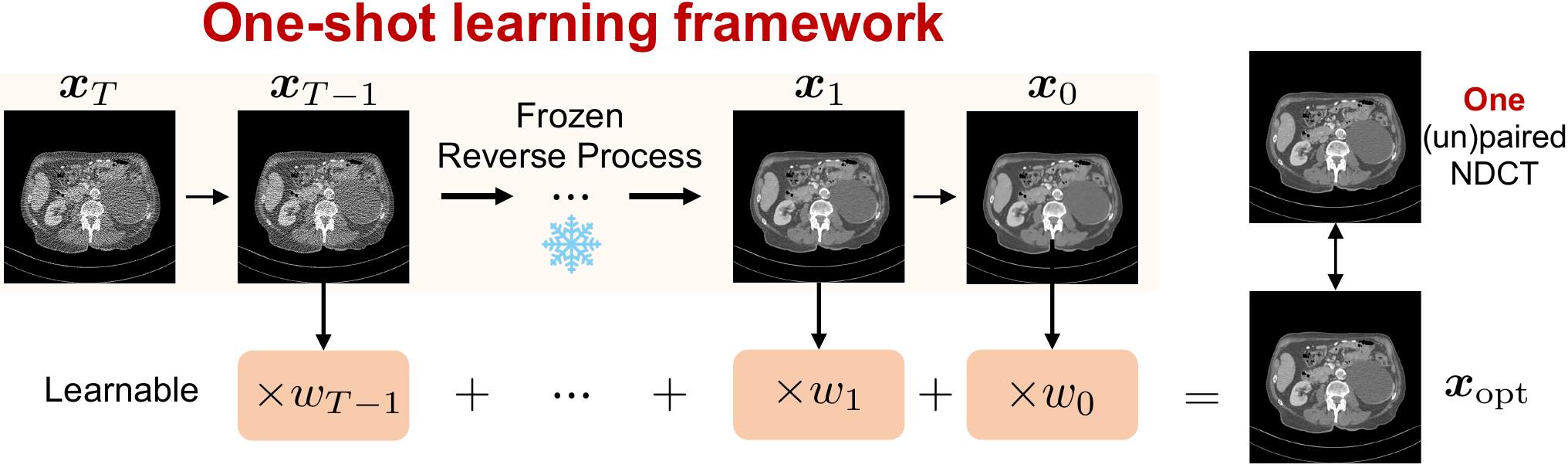}}
\caption{Framework of one-shot learning  for rapid generalization.}
\label{fig:one_shot_learning}
\end{figure}

\subsection{One-shot Learning for Rapid Generalization}\label{sec:one-shot_learning_framework}
LDCT images acquired in clinical practice are diverse due to different equipments and protocols. \emph{How to rapidly adapt one trained model to new unseen dose levels using as few as resources} is an important clinical question~\cite{xia2021ct, li2022noise, shan2019novel}. 

Here, we devise a one-shot learning (OSL) framework specifically designed for the trained \methodname with only as few as $T$ learnable parameters and enable training with one single LDCT image, as depicted in Fig.~\ref{fig:one_shot_learning}. With the introduced mean-preserving degradation operator in Eq.~\eqref{eq:proposed_degradation_operator}, the ``restoration-redegradation'' process of the \methodname is able to progressively remove the noise and artifacts from the image, yielding a series of denoised images with the same mean and varying degrees of noise. Our idea is to integrate these images to produce a visually optimal denoised image for a new, unseen dose level, which is implemented as:
\begin{align}
\vct{x}_\mathrm{opt}=\sum\nolimits_{t=0}^{T-1} w_{t} \vct{x}_{t}\  
\text {s.t.} \  \sum\nolimits_{t=0}^{T-1} w_{t}=1, \  \forall t, w_{t} \geq 0,
\label{eq:one_shot_learning}
\end{align}
where $w_{t},\  t=0, \ldots, T-1$ are the learnable weights used to synergize the images at each step, and $\vct{x}_\mathrm{opt}$ represents the optimal denoised image for the new dose level. When training this one-shot learning framework, we freeze the parameters of \networkname and only learned $w_{t}, t=0, \ldots, T-1$. Therefore, we can train the framework by dividing one single image into multiple patches. Even though the new LDCT and NDCT images are unpaired, our OSL framework would not introduce structural distortions since all $\vct{x}_{t}$ correspond to the same NDCT image.
To ensure that $\vct{x}_\mathrm{opt}$ has a better visual perception without over-smoothing, we use the perceptual loss~\cite{yang2018low} to guide the learning of those $T$ parameters.
\section{EXPERIMENTS AND RESULTS}

\subsection{Datasets}
We used four datasets in the experiments, covering different doses, centers, and objects.
\subsubsection{Mayo 2016 Dataset}
We used the ``2016 NIH-AAPM-Mayo Clinic Low-Dose CT Grand Challenge'' dataset for training and testing~\cite{chen2016open}, which contains 5,936 1mm thickness normal-dose CT slices from 10 patients. 
We randomly selected 9 patients as the training set and the remaining one as the test set.
To obtain different dose level images, a proven LDCT simulation algorithm with the widely-recognized ``Poisson+Gaussian'' noise model was used to generate low-dose projection~\cite{zeng2015simple}:
\begin{align}
\vct{p}_\mathrm{ld}=\ln \tfrac{I_{0}}{\operatorname{Poisson}(I_{0} \exp (-{\vct{p}_\mathrm{hd}}))+\operatorname{Gaussian}(0, \sigma_{e}^{2})},
\label{eq_low_dose_simulation}
\end{align}
where ${\vct{p}_\mathrm{ld}}$ and ${\vct{p}_\mathrm{hd}}$ represent the low-dose and normal-dose projections, respectively. ${I_{0}}$ is the number of incident photons, which is set to $1.5\times 10^5$. ${\sigma_{e}^{2}}$ is the variance of electronic noise, which is fixed at 10 according to~\cite{xia2021ct}. Then, the filtered back projection (FBP) algorithm was used to reconstruct images. In this experiment, we simulated 50\%, 25\%, 10\% and 5\% dose data, of which 5\% corresponds to ultra-low-dose situation~\cite{ zhao2019convolutional}. To perform a fair comparison, all deep learning methods were trained and tested on either 25\% or 5\% doses. 50\% and 25\% doses of test data were used to validate the generalization performance of our one-shot learning framework.

\subsubsection{Mayo 2020 Dataset}
To examine the generalization performance of different methods to new dose levels on the same center dataset, ``Low Dose CT Image and Projection Data'' latest released by Mayo Clinic in 2020 was used as external testing, which is named Mayo 2020 dataset~\cite{moen2021low}. This dataset contains 299 scans from two vendors, providing 25\% dose data for the head and abdomen and 10\% dose data for chest scans. We randomly selected 5 chest and 5 abdomen scans, containing 800 images for mixed dose levels testing.

\subsubsection{Piglet Dataset}
To further examine the generalization performance of different methods on a different center dataset, we also used a real piglet dataset acquired using a GE Discovery CT750 HD scanner, which contains a total of 850 CT images~\cite{yi2018sharpness}. This dataset provides 50\%, 25\%, 10\%, and 5\% dose scans corresponding to each NDCT scan. We chose two dose data 25\% and 10\% in this experiment.

\begin{figure*}[!t]
\centerline{
\includegraphics[width=0.95\textwidth]{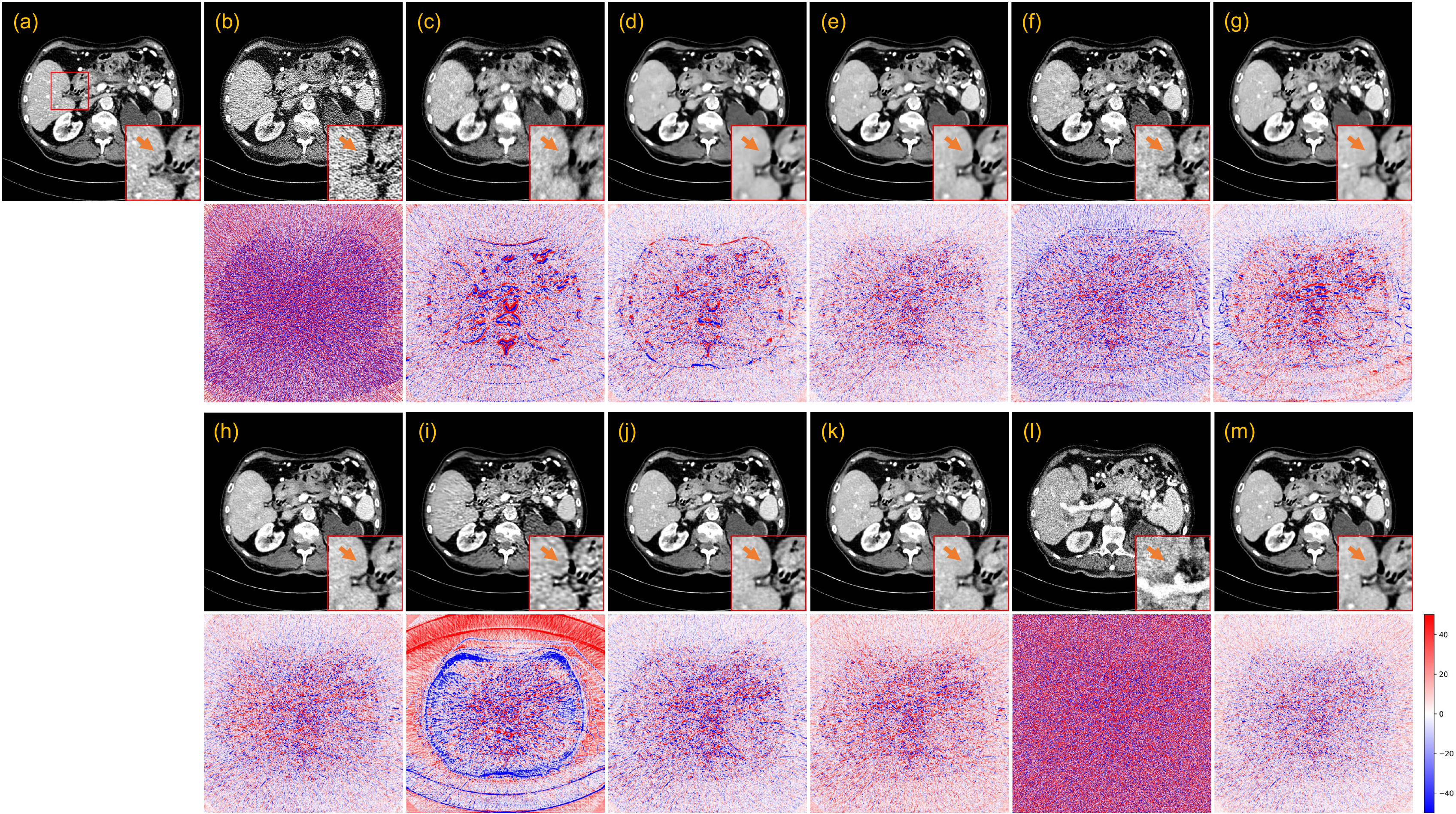}}
\caption{Qualitative results of a 25\% dose abdomen CT image from Mayo 2016 dataset. (a) NDCT image (Ground truth), (b) FBP, (c) PWLS, (d) RED-CNN, (e) PDF-RED-CNN, (f) WGAN-VGG, (g) CNCL-U-Net, (h) DU-GAN, (i) DDM$^2$, (j) IDDPM-1000, (k) IDDPM-50, (l) IDDPM-10, and (m) \methodname-10 (\textbf{ours}). The display window is [-100, 200] HU. The red ROI is zoomed in for visual comparison and the orange arrow points to one lesion.}
\label{fig:compare_results_25dose}
\end{figure*}

\begin{table*}[!t]
\centering
\caption{Quantitative results (mean$\pm$sd) on 25\% dose test data from Mayo 2016 dataset}
\label{tab:quantitative_results_on_25dose_dataset}
\begin{tabular}{lcccccc}
\shline
& PSNR(dB) & SSIM & RMSE(HU) & FSIM & VIF & NQM(dB) \\
\midrule
FBP &             34.16$\pm$1.76 & 0.7763$\pm$0.0605 & 40.0$\pm$8.0 & 0.9603$\pm$0.0108 & 0.618$\pm$0.062 & 32.40$\pm$2.18\\
PWLS &            38.26$\pm$1.29 & 0.9449$\pm$0.0091 & 24.7$\pm$3.5 & 0.9794$\pm$0.0053 & 0.633$\pm$0.052 & 29.94$\pm$2.20\\
\hline
RED-CNN &         39.29$\pm$1.53 & 0.9599$\pm$0.0106 & 22.1$\pm$4.4 & 0.9799$\pm$0.0044 & 0.574$\pm$0.033 & 29.64$\pm$1.98\\
PDF-RED-CNN &     42.94$\pm$1.32 & 0.9685$\pm$0.0103 & 14.4$\pm$2.3 & 0.9884$\pm$0.0038 & 0.667$\pm$0.049 & 35.96$\pm$1.85\\
WGAN-VGG &        40.12$\pm$0.98 & 0.9419$\pm$0.0118 & 19.9$\pm$2.3 & 0.9836$\pm$0.0031 & 0.593$\pm$0.038 & 32.97$\pm$1.39\\
CNCL-U-Net &      40.91$\pm$1.05 & 0.9598$\pm$0.0118 & 18.2$\pm$2.2 & 0.9836$\pm$0.0040 & 0.606$\pm$0.045 & 32.23$\pm$1.57\\
DU-GAN &          41.50$\pm$1.22 & 0.9591$\pm$0.0121 & 17.0$\pm$2.5 & 0.9875$\pm$0.0032 & 0.660$\pm$0.050 & 34.55$\pm$1.89\\
\hline
DDM$^2$ &         37.83$\pm$1.11 & 0.8773$\pm$0.0402 & 25.9$\pm$3.6  & 0.9766$\pm$0.0037 & 0.591$\pm$0.048  & 28.93$\pm$1.71\\
IDDPM-1000 &      41.30$\pm$1.20 & 0.9593$\pm$0.0116 & 17.4$\pm$2.5 & 0.9860$\pm$0.0036 & 0.626$\pm$0.048 & 34.20$\pm$1.85\\
IDDPM-50 &        41.49$\pm$1.18 & 0.9582$\pm$0.0122 & 17.0$\pm$2.4 & 0.9871$\pm$0.0034 & 0.634$\pm$0.047 & 34.77$\pm$1.79\\
IDDPM-10 &        33.02$\pm$1.29 & 0.6934$\pm$0.0912 & 45.2$\pm$7.0 & 0.9664$\pm$0.0075 & 0.479$\pm$0.040 & 29.79$\pm$1.29\\
\methodname-10 (\textbf{ours}) &   \textbf{43.92$\pm$1.33} & \textbf{0.9744$\pm$0.0087} & \textbf{12.9$\pm$2.1} & \textbf{0.9919$\pm$0.0026} & \textbf{0.724$\pm$0.047} & \textbf{37.84$\pm$1.76}\\
\shline 
\end{tabular}
\end{table*}

\subsubsection{Phantom Dataset}
We also used a publicly available real phantom dataset (Gammex 467 CT phantom) to examine the clinical utility of the proposed method. This dataset contains 9 different dose scans (from 33 to 499mAs) using the Thorax protocol~\cite{zhovannik2019learning}. We chose two dose data 271mAs (54.31\%) and 108mAs (21.64\%) in this experiment. For each scan, slice 10 to 21 were chosen to ensure optimal visibility of all cylindrical implants.

\begin{figure*}[!t]
\centerline{
\includegraphics[width=0.95\textwidth]{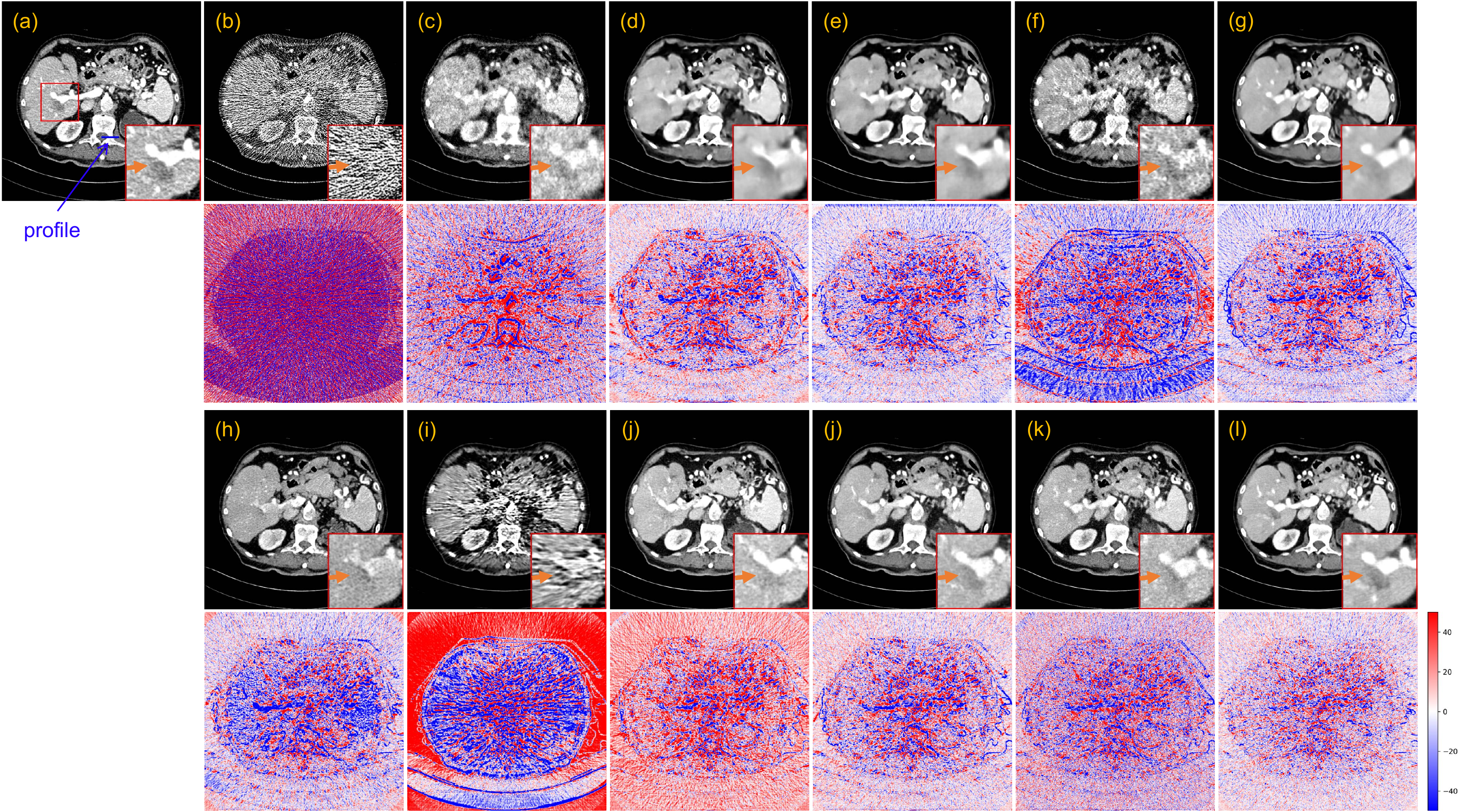}}
\caption{Qualitative results of a 5\% dose abdomen CT image from Mayo 2016 dataset. (a) NDCT image (Ground truth), (b) FBP, (c) PWLS, (d) RED-CNN, (e) PDF-RED-CNN, (f) WGAN-VGG, (g) CNCL-U-Net, (h) DU-GAN, (i) DDM$^2$, (j) IDDPM-1000, (k) IDDPM-50, (l) IDDPM-10, and (m) \methodname-10 (\textbf{ours}). The display window is [-100, 200] HU. The red ROI is zoomed in for visual comparison and the orange arrow points to one lesion.}
\label{fig:compare_results_5dose}
\end{figure*}

\begin{figure}[!t]
\centerline{\includegraphics[width=\linewidth]{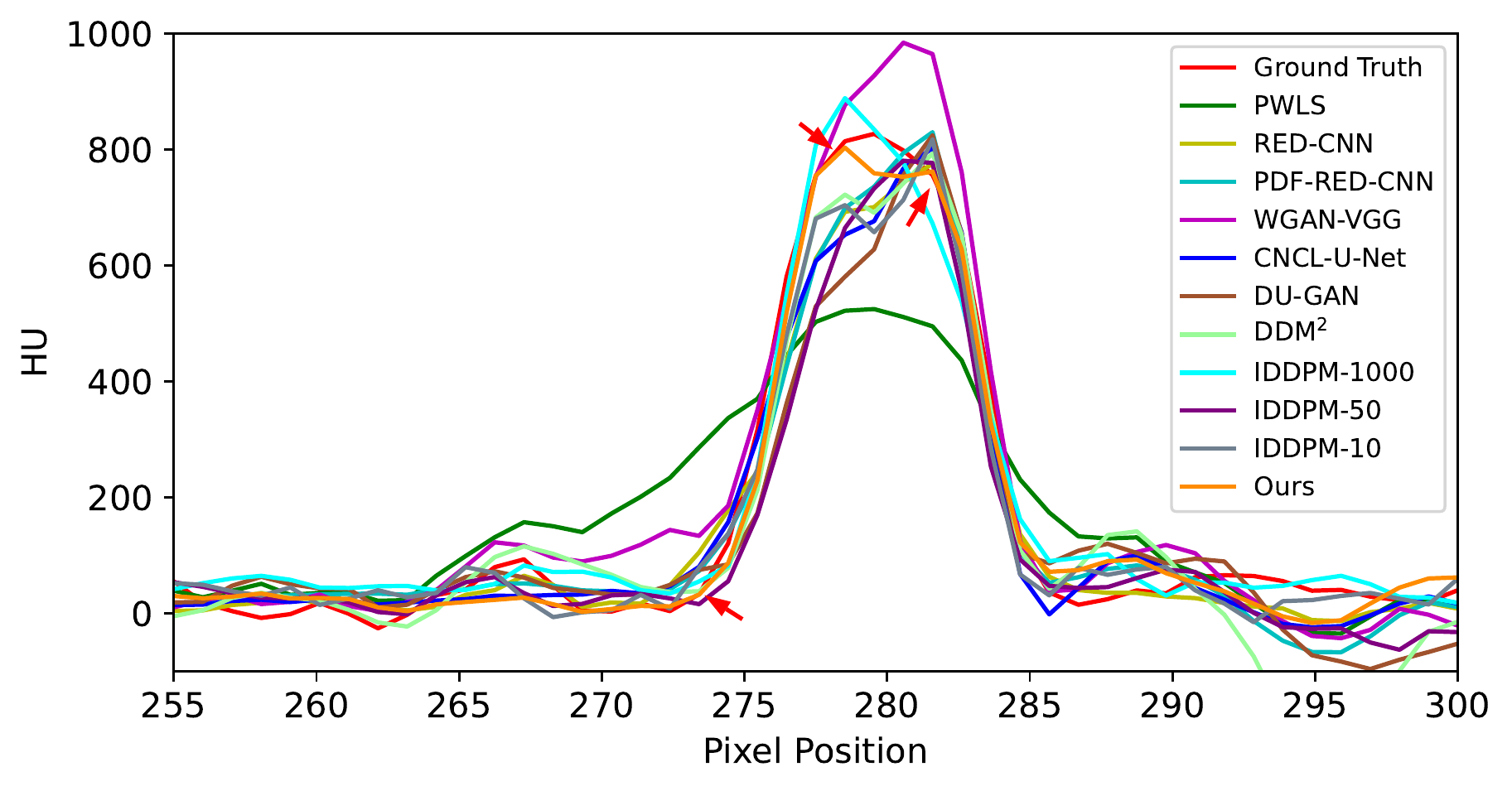}}
\caption{Profile plots of the blue line in Fig.~\ref{fig:compare_results_5dose} by different methods.}
\label{fig:profile_5dose}
\end{figure}

\subsection{Implementation Details}
Following~\cite{guo2019toward}, we used a U-Net as the backbone of the proposed \networkname, 
which consists of two downsampling blocks, one middle block, two upsampling blocks, and an output convolutional layer. 
The input to \networkname is of size ${3\times512\times 512}$ containing the contextual information of adjacent slices. We used Adam optimizer to optimize  \methodname with a learning rate of ${2\times 10^{-4}}$ and a total of 150k iterations for training.
${\alpha_{1}, \ldots, \alpha_{T}}$ were set to vary linearly from ${0.999}$ to ${0}$. We conducted data simulations based on the MIRT toolbox~\cite{shi2020review}. We implemented \methodname in PyTorch and trained it on one NVIDIA RTX 3090 GPU (24GB) with a mini-batch of size 4. For our one-shot learning framework training, we divided each image into 81 patches of size $256\times256$ with a stride of 32. The mini-batch size used for training was 8, the learning rate was set to ${2\times 10^{-3}}$, and the total training iterations were 3k. During the testing phase, we obtained the optimal denoised image by directly multiplying the images of size $512\times512$ output by \methodname with the learned weights according to Eq.~\eqref{eq:one_shot_learning}.

We compared our \methodname with four types of LDCT denoising methods, including 1) iterative reconstruction algorithm: Penalized Weighted Least Squares model (PWLS)~\cite{wang2006penalized}; 2) RED-CNN-based methods: RED-CNN~\cite{chen2017low} and PDF-RED-CNN~\cite{xia2021ct}; 3) GAN-based methods: WGAN-VGG~\cite{yang2018low}, DU-GAN~\cite{huang2021gan}, and Content-Noise Complementary Learning with U-Net (CNCL-U-Net)~\cite{geng2021content}; and 4) Diffusion-based methods: Denoising Diffusion Models for Denoising Diffusion MRI (DDM$^2$), and Improved DDPM (IDDPM)~\cite{nichol2021improved}. We set the hyperparameters of the compared DL-based methods following the original paper or official open-source codes, while all hyperparameters of PWLS adhered to the settings provided in the open-source code of~\cite{xie2017robust}. Specifically, the total iterative number of PWLS was set to 20.
We also trained an additional PDF-RED-CNN on all doses of training data from the Mayo 2016 dataset, referred to as PDF-RED-CNN\textsuperscript{$\ast$}; we adjust the 7 geometry and dose conditional parameters used in the original paper to 1 parameter, \ie, dose level. 
For the DDM$^2$ training, we replaced the slice-wise pre-trained MRI denoising model with Noise2Sim~\cite{niu2022noise}, which is a well-designed LDCT denoising model. We modified the IDDPM in reference to some works focused on developing diffusion models for LDCT image denoising tasks~\cite{gao2022cocodiff, xia2022low}.
The LDCT image was concatenated with the sampling image along the channel dimension at each time step. Then we fed the concatenated image into the network to guide IDDPM in generating the corresponding denoised image. We set $T=1000$ for IDDPM training as suggested by their original paper, and then used 1000, 50, and 10 sampling steps during inference for comparison; the resultant models were named IDDPM-1000, IDDPM-50, and IDDPM-10. Our \methodname used the same number of steps for training and inference; unless otherwise noted, $T=10$ for \methodname and the resultant model was named \methodname-10.

Three commonly-used objective image quality assessment metrics were used to quantitatively evaluate the denoising performance: peak signal-to-noise ratio (PSNR), structural similarity (SSIM) index, and root mean square error (RMSE). In addition, 
we also used three new objective IQA metrics, \ie feature similarity index (FSIM)~\cite{zhang2011fsim}, visual information fidelity (VIF)~\cite{sheikh2006image}, and noise quality metric (NQM)~\cite{damera2000image}, which have demonstrated improved alignment with subjective assessments made by radiologists on medical images~\cite{mason2019comparison}. Higher PSNR, SSIM, FSIM, VIF, and NQM and lower RMSE indicate better performance. Unless otherwise noted, all metrics were calculated based on a CT window of [-1000, 1000]HU.

\begin{table*}[!t]
\centering
\caption{Quantitative results (mean$\pm$sd) on 5\% dose test data from Mayo 2016 dataset and the computational time for denoising a single image. The reported  time for deep learning methods excludes the time required for FBP reconstruction}
\label{tab:quantitative_results_on_5dose_dataset}
\begin{tabular}{lccccccc}
\shline
& PSNR(dB) & SSIM & RMSE(HU) & FSIM & VIF & NQM(dB) & Time(s)\\
\midrule
FBP &             25.49$\pm$2.15 & 0.4310$\pm$0.0908 & 109.5$\pm$25.4 & 0.8427$\pm$0.0378 & 0.379$\pm$0.062 & 23.53$\pm$2.70 & - \\
PWLS &            34.87$\pm$0.95 & 0.8736$\pm$0.0084 & 36.3$\pm$3.9 & 0.9560$\pm$0.0073 & 0.479$\pm$0.045 & 25.58$\pm$2.03 & 3.50 \\
\hline
RED-CNN &         37.43$\pm$0.95 & 0.9363$\pm$0.0136 & 27.0$\pm$3.0 & 0.9665$\pm$0.0059 & 0.466$\pm$0.036 & 27.43$\pm$1.02 & 0.01 \\
PDF-RED-CNN &     39.25$\pm$1.22 & 0.9445$\pm$0.0147 & 22.0$\pm$3.1 & 0.9737$\pm$0.0071 & 0.527$\pm$0.049 & 29.91$\pm$1.98 & 0.01 \\
WGAN-VGG &        34.68$\pm$0.77 & 0.8821$\pm$0.0199 & 37.0$\pm$3.3 & 0.9528$\pm$0.0065 & 0.384$\pm$0.036 & 24.24$\pm$1.29 & 0.01 \\
CNCL-U-Net &      38.34$\pm$1.07 & 0.9341$\pm$0.0155 & 24.4$\pm$3.0 & 0.9684$\pm$0.0066 & 0.493$\pm$0.045 & 28.91$\pm$1.71 & 0.02\\
DU-GAN &          37.39$\pm$1.13 & 0.9287$\pm$0.0161 & 27.2$\pm$3.5 & 0.9708$\pm$0.0063 & 0.495$\pm$0.046 & 27.80$\pm$1.86 & 0.01\\
\hline
DDM$^2$ &         29.21$\pm$1.64 & 0.5908$\pm$0.0804 & \hspace{1.5mm}70.5$\pm$13.6  & 0.9206$\pm$0.0180 & 0.399$\pm$0.054  & 21.10$\pm$2.23 & 28.3\\
IDDPM-1000 &      37.31$\pm$1.14 & 0.9177$\pm$0.0263 & 27.5$\pm$3.6 & 0.9695$\pm$0.0068 & 0.482$\pm$0.046 & 28.01$\pm$2.03 & 94.2\\
IDDPM-50 &        37.95$\pm$1.28 & 0.9170$\pm$0.0362 & 25.6$\pm$3.8 & 0.9725$\pm$0.0064 & 0.500$\pm$0.047 & 29.04$\pm$2.01 & 4.67\\
IDDPM-10 &        34.80$\pm$2.52 & 0.8063$\pm$0.1125 & \hspace{1.5mm}38.1$\pm$12.8 & 0.9702$\pm$0.0056 & 0.469$\pm$0.041 & 28.20$\pm$1.86 & 0.96\\
\methodname-10 &   \textbf{40.71$\pm$1.26} & \textbf{0.9576$\pm$0.0123} & \textbf{18.6$\pm$2.7} & \textbf{0.9830$\pm$0.0048} & \textbf{0.597$\pm$0.050} & \textbf{32.36$\pm$1.95} & 0.12 \\
\shline
\end{tabular}
\end{table*}

\begin{figure}[h] 
    \centering 
    \includegraphics[width=1\linewidth]{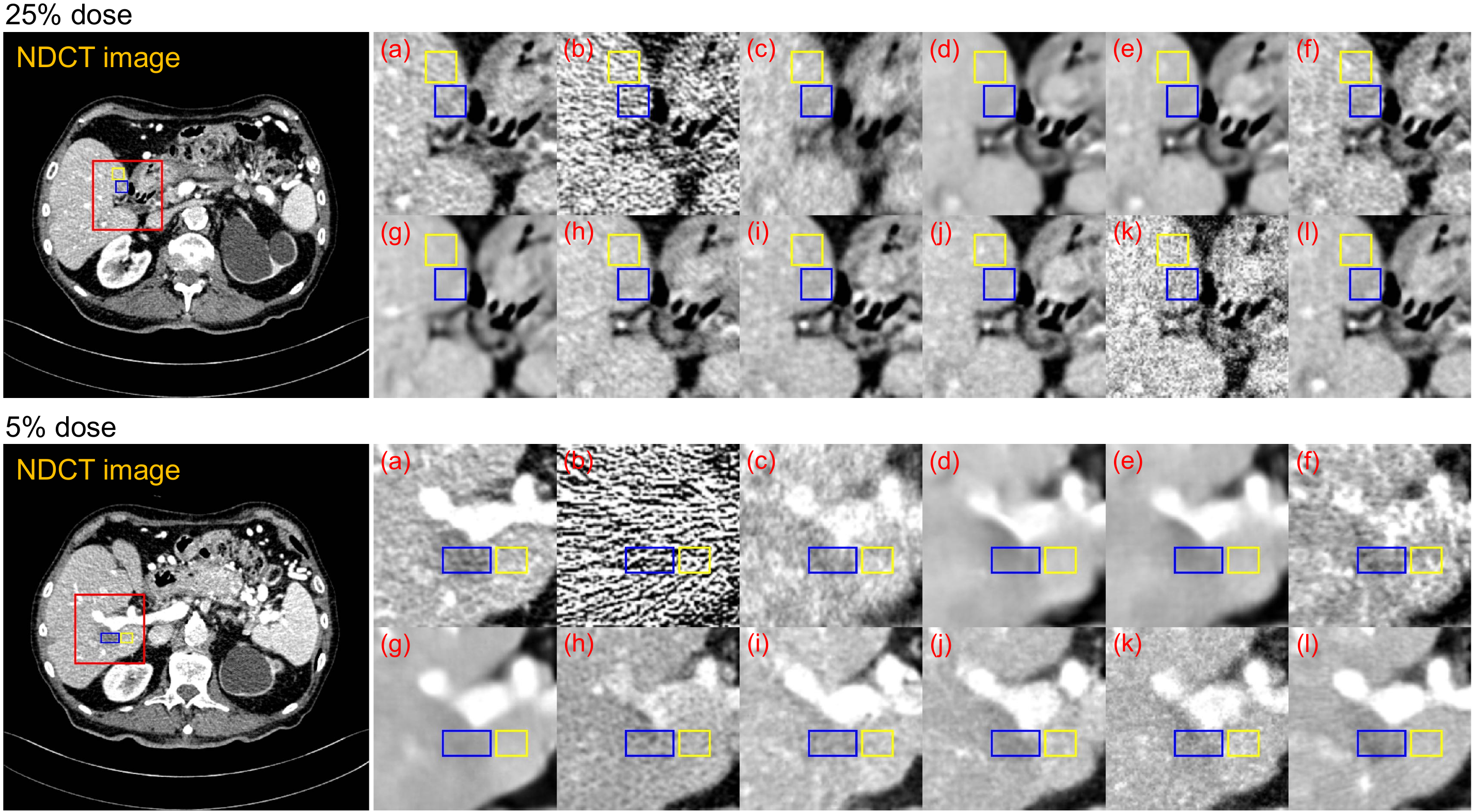} 
    \vspace{-15pt}
    \caption{Qualitative results of selected ROIs for calculating CNR. (a) NDCT image (Ground truth), (b) FBP, (c) PWLS, (d) RED-CNN, (e) PDF-RED-CNN, (f) WGAN-VGG, (g) CNCL-U-Net, (h) DU-GAN, (i) IDDPM-1000, (j) IDDPM-50, (k) IDDPM-10, and (l) \methodname-10 (\textbf{ours}). The blue ROI is extracted in the lesion region and the yellow ROI is extracted in the background. The red ROI containing the above two types of ROIs is zoomed in for visual comparison.}
    \label{fig:roi_mayo2016} 
\end{figure}

\subsection{Performance Comparison on Mayo 2016 Dataset}
In this subsection, we evaluated the denoising performance of different models on the 25\% and 5\% doses of test data from the Mayo 2016 dataset; note that all the models were also trained on the same doses.
\subsubsection{Evaluation on the 25\% dose} Fig.~\ref{fig:compare_results_25dose} presents a representative slice of 25\% dose test data denoised by different methods for visual comparison. The orange arrow indicates the location of the lesion in the red region of interest (ROI). Although the RED-CNN-based methods effectively remove noise from the LDCT image, it tends to blur fine details. Among the GAN-based methods,  WGAN-VGG introduces velvet artifacts, and DU-GAN provides textures closer to NDCT images. CNCL-U-Net preserves the most details, but its residual map shows a noticeable difference in predicting the bone edge. Among the diffusion-based models, DDM$^2$ exhibits obvious artifacts and CT number drift. We conjectured that this phenomenon may result from the fact that DDM$^2$ assumes the image noise adheres to a Gaussian distribution, which deviates from the actual noise distribution of CT images. Both in terms of texture preservation and detail retention, IDDPM and our \methodname surpass other compared methods. For the IDDPM, lowering the number of sampling steps $T$ to 50 has little impact on the denoising performance. However, when $T$ is reduced to 10, the model produces the poorest results due to much insufficient sampling. In addition, IDDPM-1000/-50 erase the critical lesion information, while our \methodname retains them well. The residual map confirms that our approach has the least prediction bias.

Table~\ref{tab:quantitative_results_on_25dose_dataset} presents the quantitative results of all methods. Our \methodname outperforms all DL-based methods and the iterative reconstruction algorithm. Notably, our method outperforms the second-best method (PDF-RED-CNN) by a large margin in terms of all metrics.

\begin{table}[!t]
\centering
\caption{Comparison of CNR and mean pixel values for lesion ROIs across different methods}
\label{tab:cnr_and_mean}
\begin{tabular}{lccccc}
\shline
 & \multicolumn{2}{c}{25\% dose} && \multicolumn{2}{c}{5\% dose}\\
\cline{2-3}\cline{5-6} 
& CNR & Mean && CNR & Mean\\
\midrule
Ground truth & 1.0456 & 106.6960 && 2.2971 & 64.6133\\
\hline
FBP &          0.3482 & 106.0759 && 0.2179 & 74.5219\\
PWLS &         1.0207 & 109.3893 && 1.3154 & 91.1295\\
\hline
RED-CNN &      \textbf{1.9308} & 110.8473 && \textbf{6.1873} & 80.9604\\
PDF-RED-CNN &  1.5268 & 110.6010 && 5.4688 & 78.2443\\
WGAN-VGG &     1.0120 & 101.5679 && 1.0363 & 71.3906\\
CNCL-U-Net &   1.0309 & 119.2151 && 4.2600 & 91.2229\\
DU-GAN &       0.9932 & 115.3734 && 1.2580 & 83.6046\\
\hline
IDDPM-1000 &   0.7855 & 112.3634 && 1.2247 & 115.3525\\
IDDPM-50 &     1.0991 & 116.8319 && 2.9835 & 99.0598\\
IDDPM-10 &     0.3222 & 115.6313 && 1.7250 & 91.9786\\
\methodname-10 (\textbf{ours}) &  1.4406 & \textbf{109.2925} && 4.5688 & \textbf{70.2043}\\
\shline 
\end{tabular}
\end{table}

\subsubsection{Evaluation on the 5\% dose} Fig.~\ref{fig:compare_results_5dose} presents the qualitative results of 5\% dose test data. In this ultra-low dose scenario, the FBP image suffers from significantly severe noise and streak artifacts due to the photon starvation effect, making it unacceptable for clinical diagnosis. The denoising performance of some denoising methods has a sharp decline. 
Fig.~\ref{fig:compare_results_5dose} shows that RED-CNN-based methods and CNCL-U-Net produce over-smoothed results. In addition, both PWLS and WGAN-VGG introduce noticeable artifacts to the denoised images. The DU-GAN obtains the best performance besides the diffusion-based methods. However, the denoising result of DU-GAN shrinks the lesion size. Other diffusion-based models, except for IDDPM-10 and DDM$^2$, consistently exhibit remarkable performance in ultra-low-dose denoising tasks,
showing great promise for LDCT denoising. Among them, our \methodname  exhibits the best denoising performance both in terms of residual maps and zoomed-in ROIs. Furthermore, Fig.~\ref{fig:profile_5dose} shows the proﬁle results of the different methods, as indicated by the blue line in the NDCT images in Fig.~\ref{fig:compare_results_5dose}. 
 The red arrow indicates that our \methodname maintains the CT number better than other methods.

Table~\ref{tab:quantitative_results_on_5dose_dataset} presents the quantitative results of 5\% dose test data. Our \methodname also surpasses all competing methods. On average our \methodname achieves around +1.46 dB  PSNR, +1.39\% SSIM, and -15.45\% RMSE over the second-best PDF-RED-CNN. In addition,  Table~\ref{tab:quantitative_results_on_5dose_dataset} also reports the computational time of denoising a single image by different methods. The inference speed of the \methodname is much faster than that of diffusion-based models, which has reached a clinically acceptable level.

In addition, we incorporated the contrast-to-noise ratio (CNR)~\cite{gutjahr2016human, park2020enhancement} to assess the detectability of low-contrast lesions in Figs.~\ref{fig:compare_results_25dose} and~\ref{fig:compare_results_5dose}.  
The higher the CNR between the lesion and the background ROIs, the increased probability of detecting low-contrast lesions.
As shown in Fig.~\ref{fig:roi_mayo2016}, we carefully selected the blue lesion ROIs and the yellow background ROIs from two slices, and the CNR values for ROIs denoised by different methods are presented in Table~\ref{tab:cnr_and_mean}. It can be observed that both RED-CNN and PDF-RED-CNN achieve the top two CNR values, while our method ranks third. 
Nonetheless, as depicted in Fig.~\ref{fig:roi_mayo2016}, both RED-CNN and PDF-RED-CNN blur the edges of lesions, which is important for doctors in staging the disease and determining its benign or malignant nature.
Considering that CT numbers are often used to differentiate healthy tissues from diseased ones in many clinical practices, we also calculated mean pixel values of lesion ROIs in Table~\ref{tab:cnr_and_mean}. Notably, our \methodname demonstrates the CT number of the lesion ROI closest to the ground truth.

\begin{figure}[!t]
\centerline{\includegraphics[width=\columnwidth]{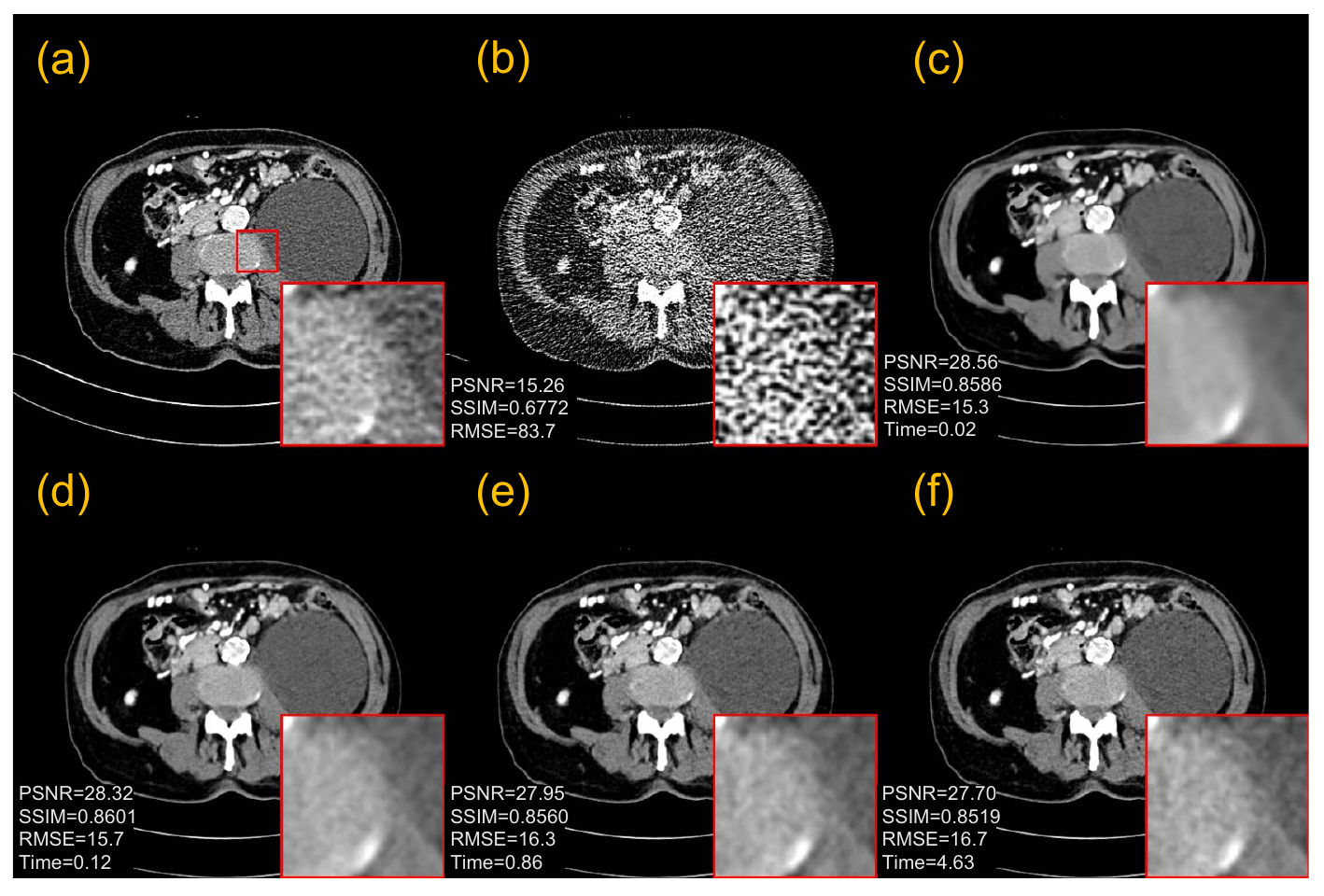}}
\caption{Ablation study on different $T$ settings for \methodname. (a) NDCT image (Ground truth), (b) FBP, (c) $T=1$, (d) $T=10$, (e) $T=50$, and (f) $T=250$. The display window is [-100, 200] HU. The tissue boundaries  in the red ROI are zoomed in for comparison. Quantitative results are provided in the lower left corner and calculated based on the display window. Time (sec.) is the average inference time of \methodname on the whole test set for different $T$ settings.}
\label{fig:ablation_study_T}
\end{figure}

\subsection{Ablation Study}\label{sec:Ablation_Study}
We conducted ablation studies to examine the effects of different ${T}$ settings and all components in \methodname. All the models were trained and tested on the 5\% dose data from Mayo 2016 dataset.

\subsubsection{Ablation on different $T$ settings}\label{sec:Ablation_on_different_T_settings} We evaluated the performance of \methodname with $T\!=\!1, 10, 50$, and $250$ for training and inference. Fig.~\ref{fig:ablation_study_T} presents the denoised images under different $T$ settings. When $T=1$, \methodname reduces to a one-step restoration, resulting in blurred edges in the denoised image. As $T$ increases, the tissue boundaries  become gradually sharper, but the inference time also increases accordingly. In addition, the prediction errors of the restoration network are also accumulated as $T$ increases. For example, when $T=1$, the PSNR and RMSE values are highest, corresponding to fewer pixel-level errors and over-smoothed images. When $T=10$, the SSIM of \methodname is the highest, and the denoised image is visually closest to the ground truth. However, setting $T\geq50$ results in a gradual decline in the quantitative performance of the \methodname. Despite the presence of our \networkname, the accumulated errors cannot be disregarded as $T$ becomes large. Therefore, considering the sharpness of the denoised images,  quantitative performance, and inference time of the method, $T=10$ is a suitable setting for our \methodname.

\begin{table}[t]
\centering
\caption{Ablation study of different modules on 5\% dose test data from Mayo 2016 dataset}
\label{tab:ablation_study_modules}
\begin{tabular}{cccccc}
\shline
\multirow{2}{*}{Warm} & \multicolumn{2}{c}{\networkname}   & PSNR & \multirow{2}{*}{SSIM} & RMSE\\
& CTX & EMM & (dB) & & (HU)\\
\midrule
\text{-} & \text{-} & \text{-} & 39.96$\pm$1.46 & 0.9501$\pm$0.0143 & 26.8$\pm$3.5\\
\checkmark & \text{-} & \text{-} & 41.59$\pm$1.57 & 0.9628$\pm$0.0123 & 22.2$\pm$3.1\\
\checkmark & \checkmark & \text{-} & 42.56$\pm$1.52 & 0.9680$\pm$0.0109 & 19.9$\pm$2.7\\
\checkmark & \checkmark & \checkmark & \textbf{43.04$\pm$1.55} & \textbf{0.9713$\pm$0.0101} & \textbf{18.8$\pm$2.7}\\
\shline
\end{tabular}
\end{table}

\begin{table}[t]
\centering
\caption{The weight distribution in the one-shot learning framework using paired or unpaired slices (pair. vs unpa.)}
\label{tab:generalization_weights}
\begin{tabular}{llc}
\shline
Dose & Type & $w_0$ $\longrightarrow$ $w_9$ \\
\hline
\multirow{2}{*}{50\%} & pair. &  0.33   0.21   0.14   0.08   0.03   0.03   0.04   0.04   0.05   0.05\\
& unpa. &  0.33   0.22   0.14   0.09   0.03   0.03   0.04   0.04   0.04   0.04\\ \hline
\multirow{2}*{25\%} & pair. & 0.44   0.29   0.12   0.04   0.02   0.01   0.02   0.02   0.02   0.02\\
&  unpa. &  0.50   0.28   0.11   0.03   0.01   0.01   0.01   0.01   0.02   0.02\\
\shline
\end{tabular}
\end{table}

\subsubsection{Ablation on different components}\label{Different_Modules} We further examined the effects of our generalized diffusion process with a mean-preserving degradation operator, the introduced contextual information, and EMM in our \networkname. For a fair comparison, we set the total sampling steps $T$ of the original cold diffusion to 10 as the baseline. For simplicity, the proposed generalized diffusion process with the LDCT image (warm state) is abbreviated as \emph{Warm} and the introduction of contextual information is abbreviated as \emph{CTX}.
Table~\ref{tab:ablation_study_modules} presents the quantitative comparison of different components, which shows that all the proposed components contribute significantly to the overall denoising performance of \methodname. On average our \methodname with all components achieves around +3.08 dB  PSNR, +2.23\%  SSIM, and -29.85\%  RMSE over the baseline.

\begin{figure}[!t]
\centerline{\includegraphics[width=\linewidth]{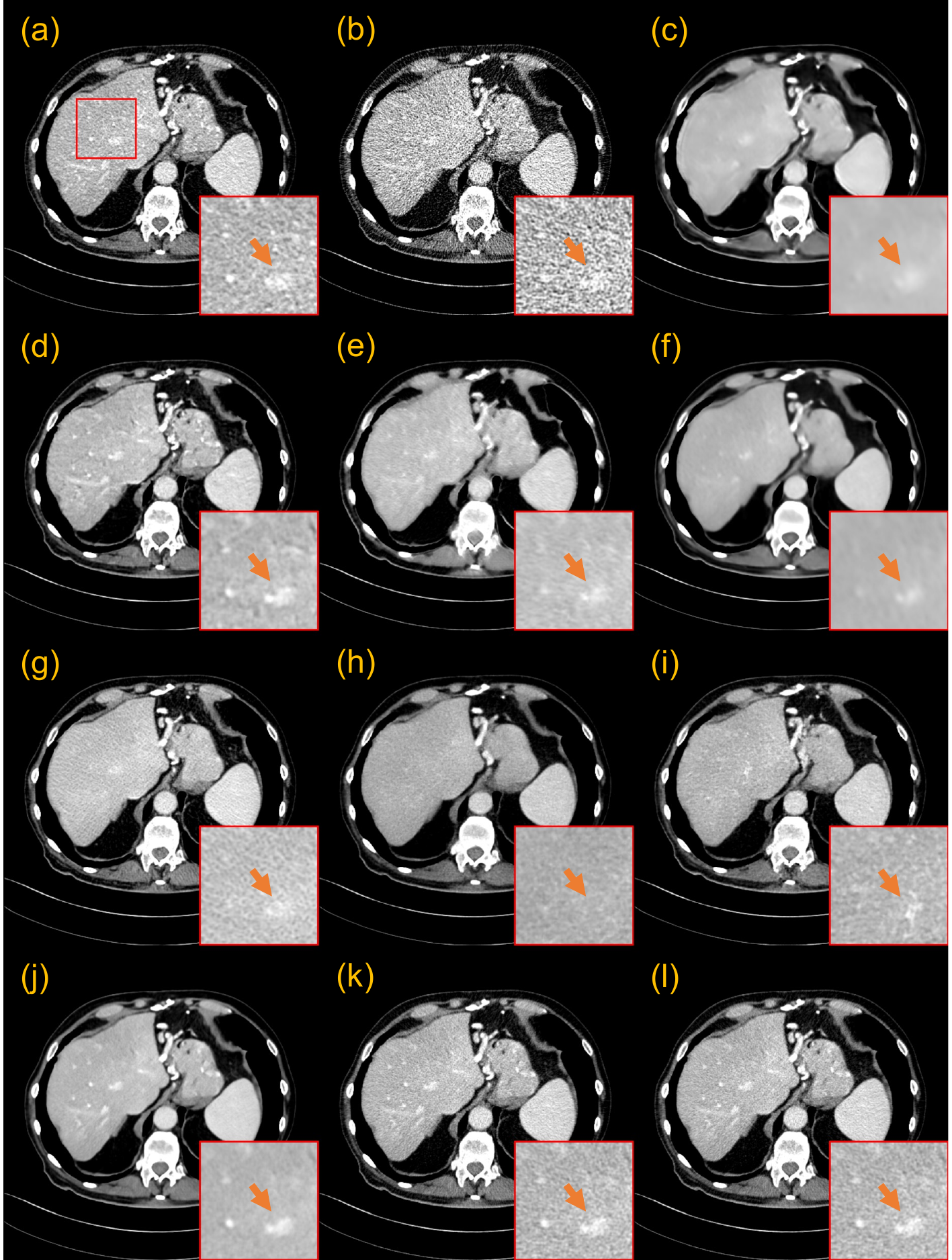}}
\caption{Qualitative results of a 50\% dose CT image denoised by different methods trained on 5\% dose training data, both training and test data from Mayo 2016 dataset. (a) NDCT image (Ground truth), (b) FBP, (c) RED-CNN, (d) PDF-RED-CNN\textsuperscript{$\ast$}, (e) WGAN-VGG, (f) CNCL-U-Net, (g) DU-GAN, (h) IDDPM-1000, (i) IDDPM-50, (j) \methodname-10 (\textbf{ours}), (k) \methodvariantp-10 (\textbf{ours}), and (l) \methodvariantu-10 (\textbf{ours}). The display window is [-100, 200] HU. The red ROI is zoomed in for visual comparison and the orange arrow points to a key detail.}
\label{fig:generalization_50dose}
\end{figure}

\begin{figure}[!t]
\centerline{\includegraphics[width=\linewidth]{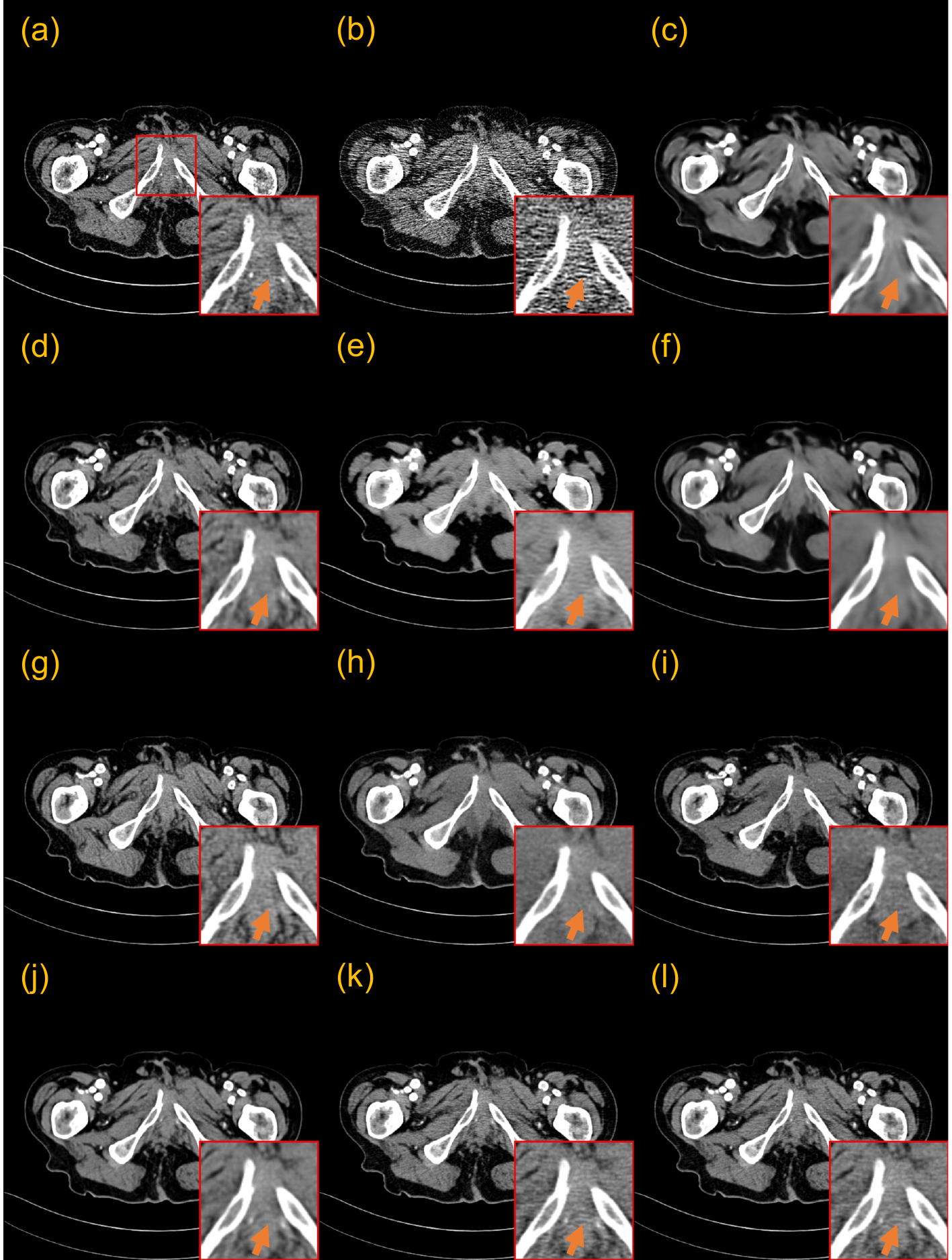}}
\caption{Qualitative results of a 25\% dose CT image denoised by different methods trained on 5\% dose training data, both training and test data from Mayo 2016 dataset. (a) NDCT image (Ground truth), (b) FBP, (c) RED-CNN, (d) PDF-RED-CNN\textsuperscript{$\ast$}, (e) WGAN-VGG, (f) CNCL-U-Net, (g) DU-GAN, (h) IDDPM-1000, (i) IDDPM-50, (j) \methodname-10 (\textbf{ours}), (k) \methodvariantp-10 (\textbf{ours}), and (l) \methodvariantu-10 (\textbf{ours}). The display window is [-100, 200] HU. The red ROI is zoomed in for visual comparison and the orange arrow points to a key detail.}
\label{fig:generalization_25dose}
\end{figure}

\subsection{One-shot Generalization to New Doses and Datasets}
We further conducted experiments on four different test datasets to evaluate the effectiveness of our one-shot learning framework, 
which can examine the generalization to 1) different doses from the same dataset (Mayo 2016 dataset), 2) different collections from the same center (Mayo 2020 dataset), 3) different species from different centers (Piglet dataset), and 4) phantom data from different centers (Phantom dataset). For all experiments in this subsection, except for PDF-RED-CNN\textsuperscript{$\ast$}, all models (including ours) were trained on the Mayo 2016 dataset with 5\% dose data. In particular, PDF-RED-CNN\textsuperscript{$\ast$} was trained using all dose data from Mayo 2016 dataset.  Additionally, the dose of test data employed in the generalization experiments varied across different datasets.
For the new dose from any dataset, we only selected one new LDCT image and one (un)paired NDCT image for training the weights of the OSL framework in Eq.~\eqref{eq:one_shot_learning}. 
If the new LDCT and NDCT images are paired, the resultant OSL model is referred to as \methodvariantp.
To ease the pairing requirement of the new training data, we also consider the unpaired scenario, in which LDCT and NDCT images were collected at different times. To achieve this, we selected a NDCT image below two slices of the corresponding LDCT image as the training label to simulate the presence of slight shifts for unpaired training; the resultant OSL model is referred to as \methodvariantu.
\subsubsection{Generalization to new dose levels on the Mayo 2016 dataset}
We examined the generalization of our \methodname to new dose levels, in which the model was trained on 5\% dose data and evaluated on 50\% and 25\% dose test data, both from the Mayo 2016 dataset. Note that comparing PDF-RED-CNN\textsuperscript{$\ast$}, \methodvariantp, and \methodvariantu to other methods is not fair because they used additional training data of 50\% and 25\% doses.

Table~\ref{tab:generalization_weights} presents the learned weights by \methodvariantp and  \methodvariantu. Weight distributions derived from paired and unpaired training are highly correlated, indicating the flexibility of the OSL framework for clinical use.

Fig.~\ref{fig:generalization_50dose} presents the qualitative results of a 50\% dose LDCT image. Both RED-CNN and CNCL-U-Net appear to smoothen the image due to the fact that the level of noise in the test image is higher than the one in the training image. While DU-GAN and IDDPM successfully preserve image texture information, certain details, such as blood vessels, are lost. Although our \methodname tends to remove more noise than necessary, it is prone to preserve critical anatomical details. We highlight our \methodvariantp and  \methodvariantu yield a visual perception effect closer to the ground truth. In contrast, PDF-RED-CNN\textsuperscript{$\ast$} may introduce some distortions. Fig.~\ref{fig:generalization_25dose} presents the qualitative results of a 25\% dose LDCT image, which can draw similar conclusions.

\begin{figure}[!t]
\centerline{\includegraphics[width=1.0\linewidth]{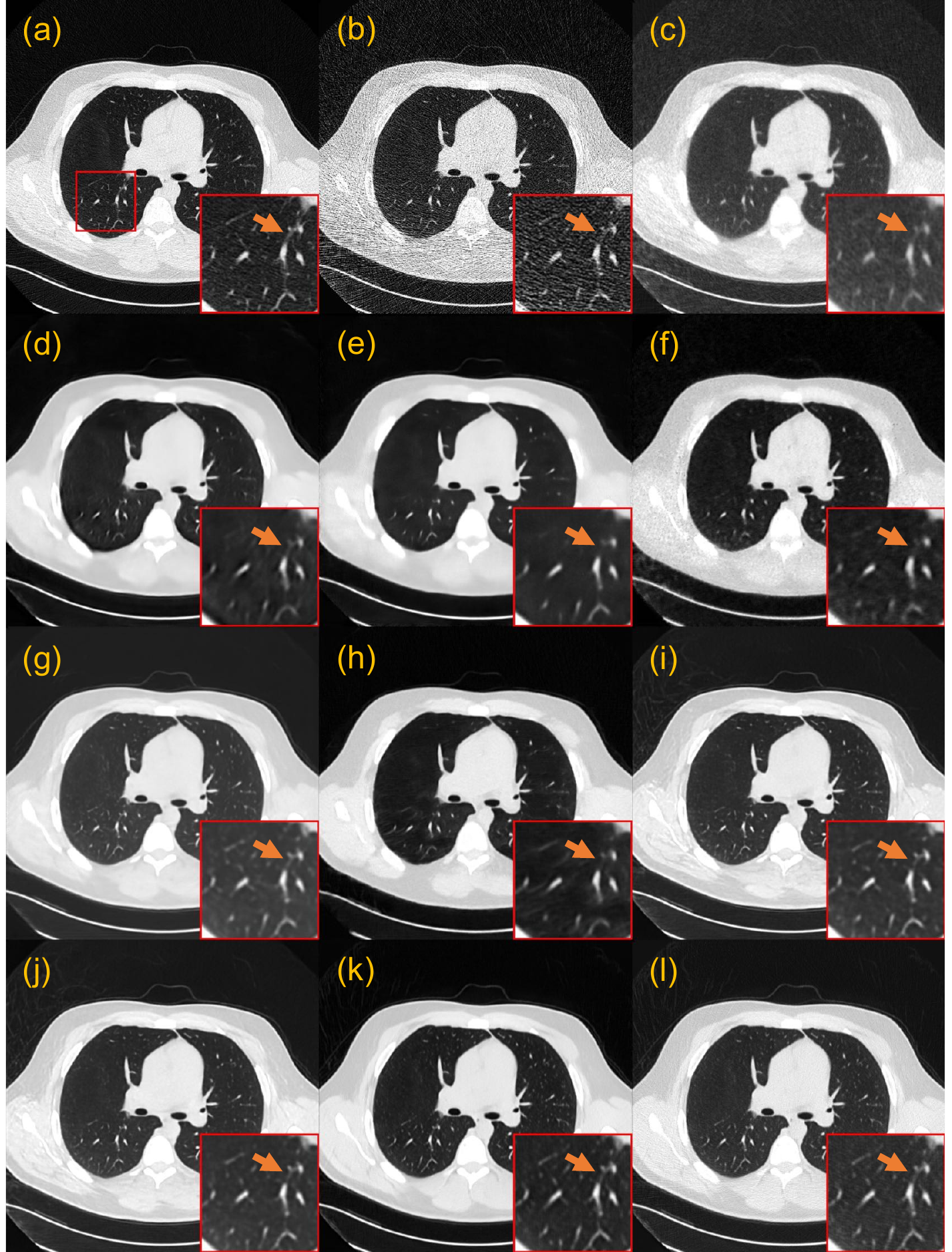}}
\caption{Qualitative results of a 10\% dose chest CT image from Mayo 2020 dataset. (a) NDCT image (Ground truth), (b) FBP, (c) PWLS, (d) RED-CNN, (e) PDF-RED-CNN\textsuperscript{$\ast$}, (f) WGAN-VGG, (g) CNCL-U-Net, (h) DU-GAN, (i) IDDPM-1000, (j) IDDPM-50, (k) \methodname-10 (\textbf{ours}), and (l) \methodvariantu-10 (\textbf{ours}). The display window is [-1350, 150] HU. The red ROI is zoomed in for visual comparison and the orange arrow points to a key detail.}
\label{fig:Mayo2020_chest}
\end{figure}

\subsubsection{Generalization to the Mayo 2020 dataset}
\label{subsubsec:mayo2020}
We further evaluated the effectiveness of the proposed \methodname and \methodvariantu on the 25\% and 10\% doses from the Mayo 2020 dataset.
To simulate the clinical application scenario as closely as possible, we chose a chest slice of 5\% dose and a abdomen slice of 25\% dose, as well as their unpaired normal-dose slices, to train two  one-shot models, respectively. By combining with these two separate one-shot models, we can quickly generalize our \methodname to mixed dose levels test data.

Fig.~\ref{fig:Mayo2020_chest} presents the denoising results of a representative chest slice from Mayo 2020 dataset. All methods remove noise to some degree. RED-CNN-based methods smoothen the images, and the numerous details in the lungs were lost in the ROI. 
PDF-RED-CNN\textsuperscript{$\ast$} leads to some detail loss partially due to the gap between our incident photon number setting for 10\% dose and the one used by Mayo Clinic.
While PWLS preserves more information compared to RED-CNN-based methods, it may not provide sufficient noise reduction. Among the GAN-based methods, CNCL-U-Net performs best but also causes the offset of the background value. 
The diffusion-based methods exhibit a good trade-off between noise suppression and image fidelity. Finally, the proposed OSL framework is proven to be instrumental in enabling \methodname to achieve textures that closely resemble the ground truth.

Table~\ref{tab:quantitative_results_on_Mayo2020_dataset} presents the quantitative results. Among the compared methods, PDF-RED-CNN\textsuperscript{$\ast$} has the highest PSNR, which benefits from the training with multiple doses of data, while IDDPM has the highest SSIM, indicating better preservation of the structural information. Surprisingly, our \methodname achieves better performance than them, even when trained only on 5\% dose data without the additional OSL framework. We conjectured that this is due to the fact that a more reasonable diffusion-based method allows the model to progressively remove noise from the image and avoid structural distortion. In addition, the contextual error modulation module makes \methodname more robust to variation of inputs.  The OSL framework further improves PSNR and SSIM, indicating that the \methodvariantu can produce more visually realistic texture and structural information by learning an optimal denoised image.

\begin{table}[!t]
\centering
\caption{Quantitative results (mean$\pm$sd) on Mayo 2020 dataset}
\label{tab:quantitative_results_on_Mayo2020_dataset}
\begin{tabular}{lccc}
\shline
& PSNR(dB) & SSIM & RMSE(HU) \\
\hline
 FBP &             29.06$\pm$8.72 & 0.6246$\pm$0.3002 & 108.7$\pm$84.2\hspace{1.5mm} \\
 PWLS &            31.96$\pm$5.29 & 0.7653$\pm$0.1947 & 60.3$\pm$32.3\\
\hline
RED-CNN &         32.57$\pm$4.83 & 0.7650$\pm$0.1918 & 54.7$\pm$27.3\\
PDF-RED-CNN\textsuperscript{$\ast$} & 33.69$\pm$6.20 &  0.7735$\pm$0.1986 & 52.2$\pm$31.0 \\
WGAN-VGG &        31.09$\pm$4.44 & 0.7475$\pm$0.1922 & 63.5$\pm$29.9\\
CNCL-U-Net &      33.15$\pm$4.87 & 0.7612$\pm$0.1954 & 51.3$\pm$25.9\\
DU-GAN &          32.73$\pm$5.42 & 0.7546$\pm$0.2030 & 55.7$\pm$30.6\\
\hline
IDDPM-1000 &  32.41$\pm$4.29 & 0.7849$\pm$0.1540 & 53.8$\pm$23.8\\
IDDPM-50 &    32.67$\pm$4.48 & 0.7887$\pm$0.1543 & 52.7$\pm$24.2\\
\methodname-10 &   34.10$\pm$5.28 & 0.7964$\pm$0.1746 & \textbf{46.8$\pm$24.5}\\
\methodvariantu-10 &     \textbf{34.13$\pm$5.37} & \textbf{0.8107$\pm$0.1626} & \textbf{46.8$\pm$24.9}\\
\shline
\end{tabular}
\end{table}

\subsubsection{Generalization to the piglet dataset}
We also examined the proposed one-shot framework in enhancing the generalization performance on the piglet dataset. We tested all the trained models on 25\% and 10\% doses of data from the piglet dataset. One slice is also selected from this piglet dataset with its (un)paired normal-dose slice to train our \methodvariantp, and \methodvariantu.

\begin{figure}[!t]
\centerline{\includegraphics[width=0.95\linewidth]{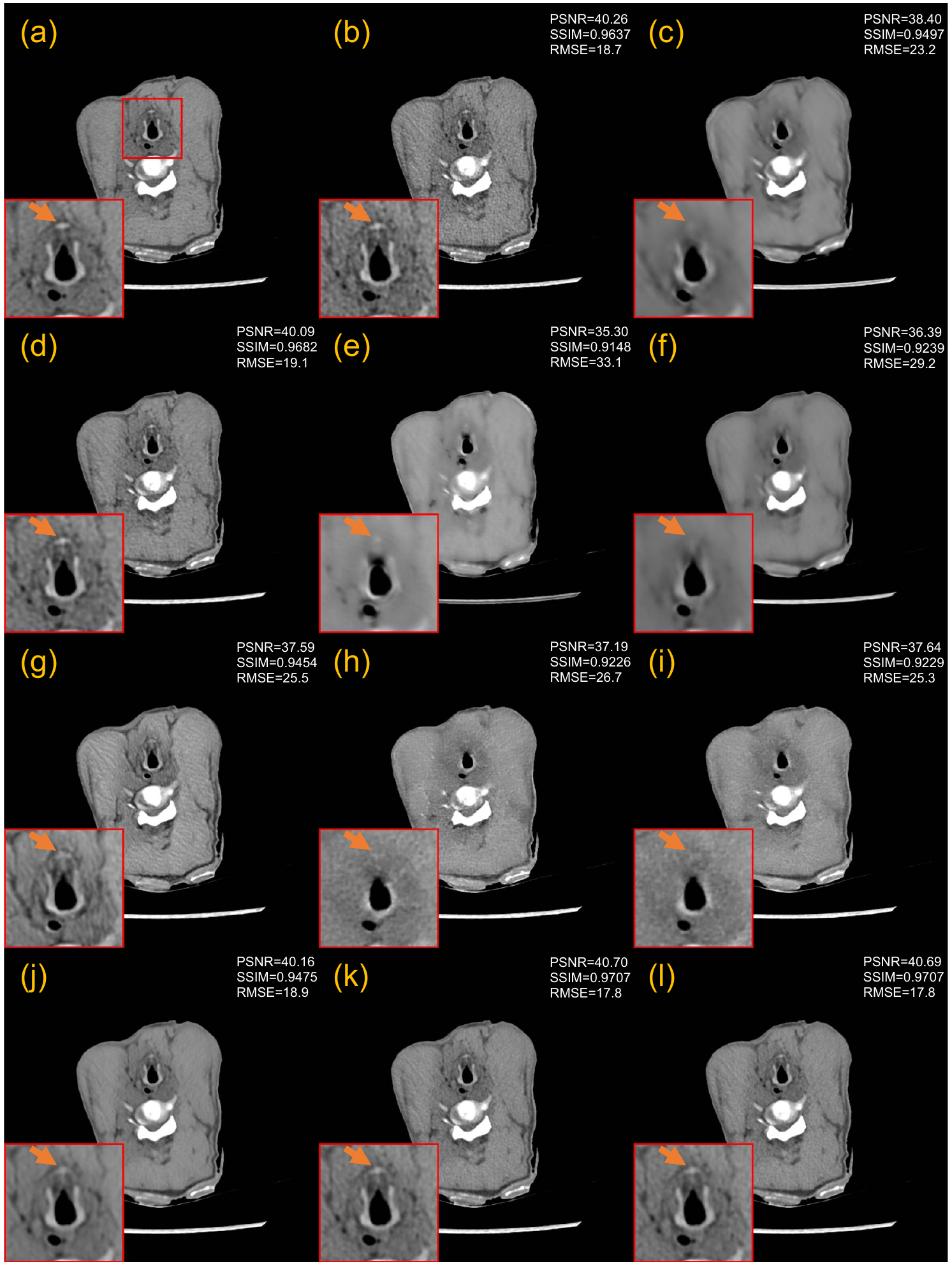}}
\caption{Qualitative results of a 25\% dose piglet CT image denoised by different methods. (a) NDCT image (Ground truth), (b) FBP, (c) RED-CNN, (d) PDF-RED-CNN\textsuperscript{$\ast$}, (e) WGAN-VGG, (f) CNCL-U-Net, (g) DU-GAN, (h) IDDPM-1000, (i) IDDPM-50, (j) \methodname-10 (\textbf{ours}), (k) \methodvariantp-10 (\textbf{ours}), and (l) \methodvariantu-10 (\textbf{ours}). The display window is [-100, 200] HU. The red ROI is zoomed in for visual comparison and the orange arrow points to a key detail.}
\label{fig:piglet_25dose}
\end{figure}

\begin{figure}[!t]
\centerline{\includegraphics[width=0.95\linewidth]{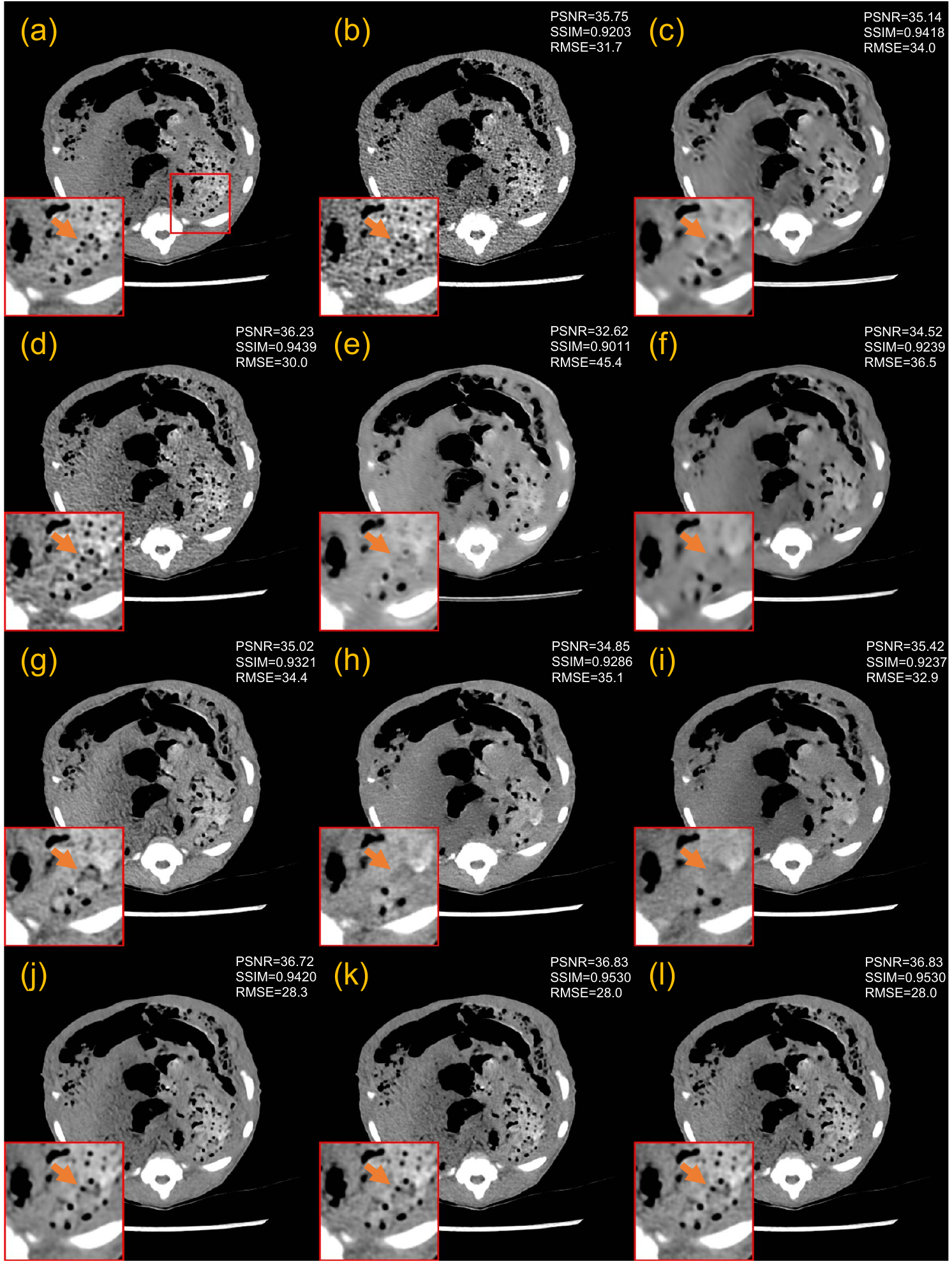}}
\caption{Qualitative results of a 10\% dose piglet CT image denoised by different methods. (a) NDCT image (Ground truth), (b) FBP, (c) RED-CNN, (d) PDF-RED-CNN\textsuperscript{$\ast$}, (e) WGAN-VGG, (f) CNCL-U-Net, (g) DU-GAN, (h) IDDPM-1000, (i) IDDPM-50, (j) \methodname-10 (\textbf{ours}), (k) \methodvariantp-10 (\textbf{ours}), and (l) \methodvariantu-10 (\textbf{ours}). The display window is [-100, 200] HU. The red ROI is zoomed in for visual comparison and the orange arrow points to a key detail.}
\label{fig:piglet_10dose}
\end{figure}

Figs.~\ref{fig:piglet_25dose} and~\ref{fig:piglet_10dose} show the denoising results of two representative slices of 25\% and 10\% doses. In the 10\% dose scenario, RED-CNN, WGAN-VGG, and CNCL-U-Net severely blur the denoised image. While DU-GAN and IDDPM manage to avoid global smoothing, they still result in the loss of crucial local details. PDF-RED-CNN\textsuperscript{$\ast$} and \methodname successfully preserve fine details. When integrated with the OSL framework, the denoised images from \methodvariantp and \methodvariantu exhibit textures closest to the NDCT image. \methodvariantp and \methodvariantu also improve the quantitative performances over \methodname, and outperform other methods. In the 25\% dose scenario, surprisingly, the quantitative metrics of the remaining methods except \methodvariantp and \methodvariantu are even worse than that of FBP. This observation highlights the limitations of DL-based models when generalizing across multi-center, multi-species CT data. However, the denoised images by \methodvariantp and \methodvariantu within our OSL framework consistently demonstrate superior visual quality and quantitative performance.

\subsubsection{Generalization to the phantom dataset}
\begin{figure}[!t]
\centerline{\includegraphics[width=0.95\linewidth]{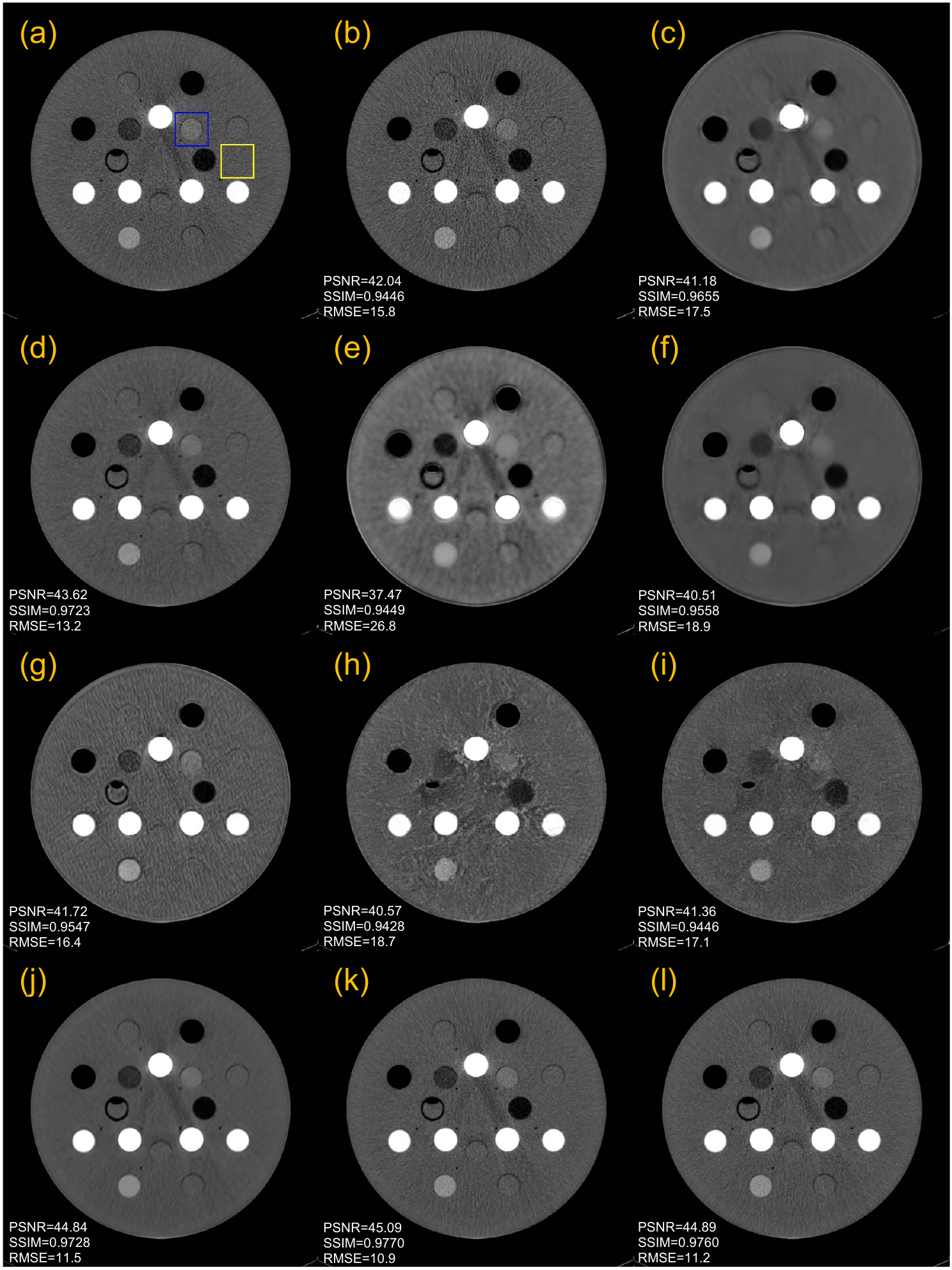}}
\caption{Qualitative results of a 54.31\% dose phantom CT image denoised by different methods. (a) NDCT image (Ground truth), (b) FBP, (c) RED-CNN, (d) PDF-RED-CNN\textsuperscript{$\ast$}, (e) WGAN-VGG, (f) CNCL-U-Net, (g) DU-GAN, (h) IDDPM-1000, (i) IDDPM-50, (j) \methodname-10 (\textbf{ours}), (k) \methodvariantp-10 (\textbf{ours}), and (l) \methodvariantu-10 (\textbf{ours}). The display window is [-100, 200] HU. The blue signal-present ROI and yellow signal-absent background ROI are selected as the input to CHO for signal detection.}
\label{fig:phantom_271}
\end{figure}

\begin{figure}[!t]
\centerline{\includegraphics[width=0.95\linewidth]{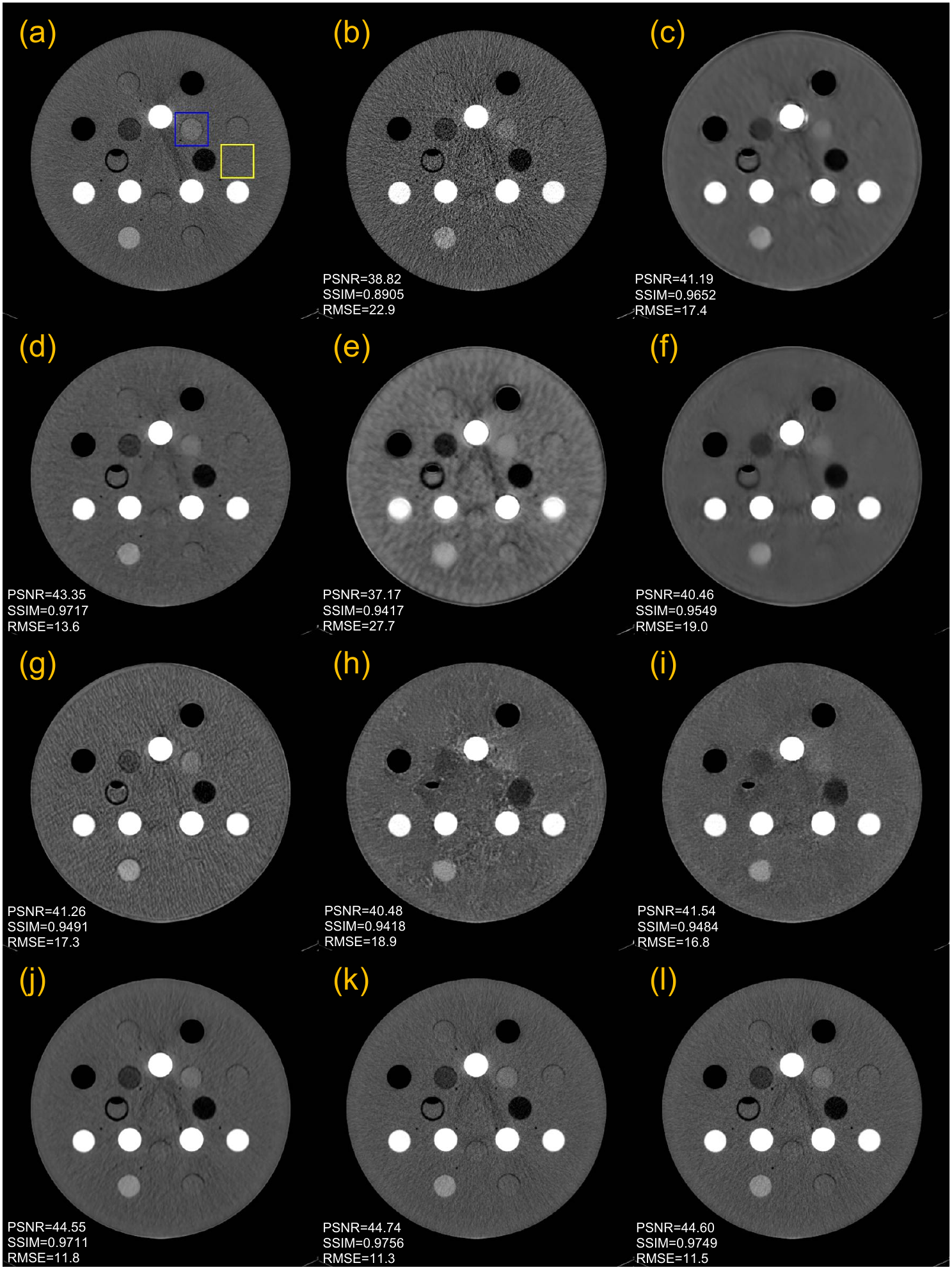}}
\caption{Qualitative results of a 21.64\% dose phantom CT image denoised by different methods. (a) NDCT image (Ground truth), (b) FBP, (c) RED-CNN, (d) PDF-RED-CNN\textsuperscript{$\ast$}, (e) WGAN-VGG, (f) CNCL-U-Net, (g) DU-GAN, (h) IDDPM-1000, (i) IDDPM-50, (j) \methodname-10 (\textbf{ours}), (k) \methodvariantp-10 (\textbf{ours}), and (l) \methodvariantu-10 (\textbf{ours}). The display window is [-100, 200] HU. The blue signal-present ROI and yellow signal-absent background ROI are selected as the input to CHO for signal detection.}
\label{fig:phantom_108}
\end{figure}

To further examine the clinical utility of the proposed \methodname, \methodvariantp, and \methodvariantu, we conducted generalization experiments on the phantom dataset and chose a channelized Hotelling observer (CHO) to quantitatively assess the performance of different methods on the low-contrast signal detection task. The utility of CHO has been explored in assessing the low-contrast detection task performance of different CT image reconstruction algorithms, demonstrating a high correlation with human observers~\cite{van2011task, wunderlich2014exact, yu2017correlation}.
The signal-to-noise ratio (SNR) is used to measure the detection performance of CHO using images denoised by different methods as input; the larger, the better.
As shown in Figs.~\ref{fig:phantom_271} and~\ref{fig:phantom_108}, we fed the blue signal-present ROIs and the yellow signal-absent background ROIs into CHO for signal detection. All trained models are tested on 54.31\% and 21.64\% doses of data from this phantom dataset. For PDF-RED-CNN\textsuperscript{$\ast$}, we utilized the nearest corresponding doses (25\% and 50\%) as conditional inputs within the parameter-dependent framework.

\begin{table}[!t]
\centering
\caption{SNR of channelized Hotelling observers using images denoised by different methods as input}
\label{tab:snr_of_cho}
\begin{tabular}{lrr}
\shline
& 271 mAs (54.31\%) & 108 mAs (21.64\%)\\
\midrule
NDCT &                  10.92 & 10.92 \\
FBP &                    9.09 & 7.71 \\
\hline
RED-CNN &                5.72 & 4.49 \\
PDF-RED-CNN\textsuperscript{$\ast$} &            7.95 & 6.70 \\
WGAN-VGG &               7.51 & 4.95 \\
CNCL-U-Net &             8.09 & 5.68\\
DU-GAN &                 7.45 & 7.71\\
\hline
IDDPM-1000 &             4.37 & 4.10\\
IDDPM-50 &               2.58 & 2.49\\
\methodname-10 &         8.24 & 8.03\\
\methodvariantp-10 &     \textbf{11.48} & \textbf{9.40}\\
\methodvariantu-10 &     11.39 & 9.36\\
\shline 
\end{tabular}
\end{table}

Figs.~\ref{fig:phantom_271} and~\ref{fig:phantom_108} present a representative slice of 54.31\% and 21.64\% doses denoised by different methods. RED-CNN and CNCL-U-Net blur the edges of cylindrical implants, while WGAN-VGG and IDDPM introduce obvious artifacts. DU-GAN exhibits inadequate noise suppression in denoised images. Although PDF-RED-CNN\textsuperscript{$\ast$} outperforms other compared methods, it falls short in preserving edge details in the 21.64\% dose scenario compared to our \methodname. The proposed OSL framework aids our model in producing background textures that are close to NDCT images, resulting in more favorable visual results.

Table~\ref{tab:snr_of_cho} presents the quantitative task-specific performance of different methods. The performance of methods besides \methodname is even inferior to that of FBP, which may be attributed to the shift in distribution between the test data (phantom) and the training data (patient). Consistent with the qualitative analysis, the performance of our \methodname is still optimal. When integrated with the OSL framework, the task-specific performance of our model was further enhanced, even surpassing NDCT in the 54.31\% dose scenario.

\section{DISCUSSION}

\subsection{Discussion on the Benefits of \methodname}
The experimental results show that our \methodname outperforms other competing models in terms of quantitative, qualitative, and task-specific performance, demonstrating the potential of generalized diffusion models for LDCT denoising. It should be noted that even with 1,000-step sampling, the IDDPM modified for LDCT denoising task performs poorly than our \methodname.
Here, we discuss the advantages of the proposed \methodname compared to other diffusion-based LDCT denoising methods from the following three aspects.
1) \emph{Benefits from the deterministic sampling process:}Recently, several LDCT denoising methods based on diffusion models have been proposed~\cite{gao2022cocodiff, xia2022low}. However, most of them follow the classical Gaussian diffusion model framework, transforming the LDCT image denoising task into a conditional generation task. The sampling process of Gaussian diffusion models can be unified into a stochastic differential equation (SDE) to guarantee good target distribution coverage~\cite{song2020score}. However, it is important to note that the image denoising task differs from the image generation task, as it solely requires establishing a one-to-one map between the denoised image and the clean image. 
In this work, we built a deterministic sampling process based on a generalized diffusion model, which not only speeds up the sampling process but also benefits image denoising tasks. 
2) \emph{Benefits from the proposed mean-preserving degradation operator:} 
Fig.~\ref{fig:different_diffusion_process} shows that the commonly-used degradation operator in Gaussian diffusion models deviates from the physical process of CT image degradation as the dose decreases. The proposed mean-preserving degradation operator not only introduces LDCT image-specific noise and artifacts into each time step of the diffusion process, but also ensures that for each pixel in the intermediate image, its CT number should have the same mean value as the one in the NDCT image.
3) \emph{Benefits from the proposed \networkname:} The issue of error accumulation during the sampling process of diffusion models is more critical for image denoising tasks due to the pursuit
of pixel-level accuracy. In this work, we proposed a novel restoration network \networkname, which can leverage contextual information to constrain the sampling process from structural distortion and modulate time step embedding features for better alignment with the input at the next time step.

\subsection{Discussion on the One-shot Learning Framework}
We also evaluated the proposed one-shot learning framework on four datasets and extensive experimental results confirmed the strong generalization over existing methods with as few as resources. 
Here, we considered 5\% to be an ultra-low-dose situation, and the actual dose used in the majority of clinical scans exceeds 5\%. However, when the dose of test data is lower than one of the training data, the output of \methodname without OSL after 10 sampling steps is already the optimal denoised result we can obtain. Hence, further research is required to explore how our \methodname can generalize from high-dose available training data to low-dose test data.
Recent works such as GMM-U-Net~\cite{li2022noise} and PDF~\cite{xia2021ct} have also focused on the generalization issue, and the main differences between our work and them are as follows. 1) They were based on previous deep learning frameworks, which are limited in LDCT imaging quality. Our work explored how to enhance the robustness of diffusion models in dealing with intricate LDCT denoising scenarios at the first time. 2) PDF takes the input of dose, imaging geometry, and other information as conditions during the training phase to enhance its robustness in reconstructing various test data. However, its performance on test data with geometry and dose levels not encountered during training will be compromised. GMM-U-Net adopted a Gaussian mixture model (GMM) to characterize the noise distribution of LDCT images, necessitating the empirical selection of different numbers of Gaussian models and loss function weights for diverse datasets. Compared with them, we only used one new LDCT image and one (un)paired NDCT image for our one-shot learning framework training, eliminating the need for empirical hyperparameter tuning on the test data, thus enabling rapid adaptation to new unseen dose levels. 3) We validated the potential of \methodname on the highly demanding task of ultra-low-dose imaging. In contrast, the doses of test data they used were higher than our ultra-low-dose level.

\subsection{Discussion on the Limitations}
We acknowledge some limitations in this work. First, in practical scenarios, since the mean of noise in LDCT images may not be exactly 0, the proposed degradation operator is not a strictly ``mean-preserving degradation operator''.  
The causes of noise in LDCT images are notably intricate and can be influenced by a multitude of factors, including scattering, beam hardening, patient motion, metal implants, etc~\cite{kalisz2016artifacts}. Therefore, further improvements of the degradation operator are required to better approximate the actual LDCT physical degradation process.
Second, although the sampling speed of our \methodname is considerably faster than that of DDM$^2$ and IDDPM, its inference time is still 10 times that of RED-CNN-based and GAN-based models. Several recent techniques aimed at accelerating diffusion models could potentially be integrated into our \methodname to further reduce sampling steps. Examples include latent space sampling technology employed by stable diffusion~\cite{rombach2022high} and the knowledge distillation technology utilized by consistency models~\cite{song2023consistency}. However, a reduction in sampling steps implies that the one-shot learning framework would utilize fewer intermediate images, necessitating a trade-off between generalization performance and inference time.
Third, since the test data comprises only two lesions, conducting a task-driven reader study does not yield statistically significant results. We intend to incorporate a larger test patient dataset in future studies to perform a multi-reader, multi-case study (MRMC) study.
\section{CONCLUSION}
In this work, we proposed a novel contextual error-modulated generalized diffusion model for LDCT image denoising, termed \methodname. The presented \methodname utilizes (i) the LDCT images as the informative endpoint of the diffusion process, (ii) the introduced mean-preserving degradation operator to mimic the physical process of CT degradation, 
(iii) the proposed restoration network \networkname to alleviate the accumulated error and misalignment, and (iv) the devised one-shot learning framework to boost the generalization.  
Experimental results demonstrate the effectiveness of \methodname, especially in the ultra-low-dose case. 
Remarkably, our \methodname model only requires 10 sampling steps, making it much faster than classical diffusion models for clinical use.



\begin{thebibliography}{10}
\providecommand{\url}[1]{#1}
\csname url@samestyle\endcsname
\providecommand{\newblock}{\relax}
\providecommand{\bibinfo}[2]{#2}
\providecommand{\BIBentrySTDinterwordspacing}{\spaceskip=0pt\relax}
\providecommand{\BIBentryALTinterwordstretchfactor}{4}
\providecommand{\BIBentryALTinterwordspacing}{\spaceskip=\fontdimen2\font plus
\BIBentryALTinterwordstretchfactor\fontdimen3\font minus
  \fontdimen4\font\relax}
\providecommand{\BIBforeignlanguage}[2]{{%
\expandafter\ifx\csname l@#1\endcsname\relax
\typeout{** WARNING: IEEEtran.bst: No hyphenation pattern has been}%
\typeout{** loaded for the language `#1'. Using the pattern for}%
\typeout{** the default language instead.}%
\else
\language=\csname l@#1\endcsname
\fi
#2}}
\providecommand{\BIBdecl}{\relax}
\BIBdecl

\bibitem{smith2009radiation}
R.~Smith-Bindman \emph{et~al.}, ``Radiation dose associated with common
  computed tomography examinations and the associated lifetime attributable
  risk of cancer,'' \emph{Arch. Intern. Med.}, vol. 169, pp. 2078--2086, 2009.

\bibitem{sodickson2009recurrent}
A.~Sodickson \emph{et~al.}, ``Recurrent {CT}, cumulative radiation exposure,
  and associated radiation-induced cancer risks from {CT} of adults,''
  \emph{Radiology}, vol. 251, pp. 175--184, 2009.

\bibitem{xie2017robust}
Q.~Xie \emph{et~al.}, ``Robust low-dose {CT} sinogram preprocessing via
  exploiting noise-generating mechanism,'' \emph{IEEE Trans. Med. Imag.},
  vol.~36, no.~12, pp. 2487--2498, 2017.

\bibitem{wang2006penalized}
J.~Wang, T.~Li, H.~Lu, and Z.~Liang, ``Penalized weighted least-squares
  approach to sinogram noise reduction and image reconstruction for low-dose
  {X}-ray computed tomography,'' \emph{IEEE Trans. Med. Imag.}, vol.~25,
  no.~10, pp. 1272--1283, 2006.

\bibitem{shan2018d}
H.~Shan \emph{et~al.}, ``3-{D} convolutional encoder-decoder network for
  low-dose {CT} via transfer learning from a 2-{D} trained network,''
  \emph{IEEE Trans. Med. Imag.}, vol.~37, no.~6, pp. 1522--1534, 2018.

\bibitem{aharon2006k}
M.~Aharon, M.~Elad, and A.~Bruckstein, ``{K-SVD}: An algorithm for designing
  overcomplete dictionaries for sparse representation,'' \emph{IEEE Trans.
  Signal Process.}, vol.~54, no.~11, pp. 4311--4322, 2006.

\bibitem{feruglio2010block}
P.~F. Feruglio, C.~Vinegoni, J.~Gros, A.~Sbarbati, and R.~Weissleder, ``Block
  matching {3D} random noise filtering for absorption optical projection
  tomography,'' \emph{Phys. Med. Biol.}, vol.~55, no.~18, p. 5401, 2010.

\bibitem{chen2013improving}
Y.~Chen \emph{et~al.}, ``Improving abdomen tumor low-dose {CT} images using a
  fast dictionary learning based processing,'' \emph{Phys. Med. Biol.},
  vol.~58, no.~16, p. 5803, 2013.

\bibitem{ma2011low}
J.~Ma \emph{et~al.}, ``Low-dose computed tomography image restoration using
  previous normal-dose scan,'' \emph{Med. Phys.}, vol.~38, pp. 5713--5731,
  2011.

\bibitem{li2014adaptive}
Z.~Li \emph{et~al.}, ``Adaptive nonlocal means filtering based on local noise
  level for {CT} denoising,'' \emph{Med. Phys.}, vol.~41, no.~1, p. 011908,
  2014.

\bibitem{sheng2014denoised}
K.~Sheng, S.~Gou, J.~Wu, and S.~X. Qi, ``Denoised and texture enhanced {MVCT}
  to improve soft tissue conspicuity,'' \emph{Med. Phys.}, vol.~41, no.~10, p.
  101916, 2014.

\bibitem{shan2019competitive}
H.~Shan \emph{et~al.}, ``Competitive performance of a modularized deep neural
  network compared to commercial algorithms for low-dose {CT} image
  reconstruction,'' \emph{Nat. Mach. Intell.}, vol.~1, pp. 269--276, 2019.

\bibitem{wang2020deep}
G.~Wang, J.~C. Ye, and B.~De~Man, ``Deep learning for tomographic image
  reconstruction,'' \emph{Nat. Mach. Intell.}, vol.~2, no.~12, pp. 737--748,
  2020.

\bibitem{chen2017low}
H.~Chen \emph{et~al.}, ``Low-dose {CT} with a residual encoder-decoder
  convolutional neural network,'' \emph{IEEE Trans. Med. Imag.}, vol.~36, pp.
  2524--2535, 2017.

\bibitem{xia2021ct}
W.~Xia \emph{et~al.}, ``{CT} reconstruction with {PDF}: {Parameter}-dependent
  framework for data from multiple geometries and dose levels,'' \emph{IEEE
  Trans. Med. Imag.}, vol.~40, no.~11, pp. 3065--3076, 2021.

\bibitem{kim2019performance}
B.~Kim, M.~Han, H.~Shim, and J.~Baek, ``A performance comparison of
  convolutional neural network-based image denoising methods: {The} effect of
  loss functions on low-dose {CT} images,'' \emph{Med. Phys.}, vol.~46, pp.
  3906--3923, 2019.

\bibitem{nagare2021bias}
M.~Nagare, R.~Melnyk, O.~Rahman, K.~D. Sauer, and C.~A. Bouman, ``A
  bias-reducing loss function for {CT} image denoising,'' in \emph{Proc. IEEE
  Int. Conf. Acoust. Speech Signal Process.}, 2021, pp. 1175--1179.

\bibitem{kang2019cycle}
E.~Kang, H.~J. Koo, D.~H. Yang, J.~B. Seo, and J.~C. Ye, ``Cycle consistent
  adversarial denoising network for multiphase coronary {CT} angiography,''
  \emph{Med. Phys.}, vol.~46, pp. 550--562, 2019.

\bibitem{yang2018low}
Q.~Yang \emph{et~al.}, ``Low-dose {CT} image denoising using a generative
  adversarial network with {Wasserstein} distance and perceptual loss,''
  \emph{IEEE Trans. Med. Imag.}, vol.~37, pp. 1348--1357, 2018.

\bibitem{huang2021gan}
Z.~Huang, J.~Zhang, Y.~Zhang, and H.~Shan, ``{DU}-{GAN}: {Generative}
  adversarial networks with dual-domain {U}-{Net} based discriminators for
  low-dose {CT} denoising,'' \emph{IEEE Trans. Instrum. Meas.}, vol.~71, pp.
  1--12, 2021.

\bibitem{zhao2018bias}
S.~Zhao, H.~Ren, A.~Yuan, J.~Song, N.~Goodman, and S.~Ermon, ``Bias and
  generalization in deep generative models: {An} empirical study,'' in
  \emph{Proc. Adv. Neural Inf. Process. Syst.}, vol.~31, 2018.

\bibitem{sohl2015deep}
J.~Sohl-Dickstein, E.~Weiss, N.~Maheswaranathan, and S.~Ganguli, ``Deep
  unsupervised learning using nonequilibrium thermodynamics,'' in \emph{Proc.
  Int. Conf. Mach. Learn.}, 2015, pp. 2256--2265.

\bibitem{ho2020denoising}
J.~Ho, A.~Jain, and P.~Abbeel, ``Denoising diffusion probabilistic models,'' in
  \emph{Proc. Adv. Neural Inf. Process. Syst.}, vol.~33, 2020, pp. 6840--6851.

\bibitem{song2020denoising}
J.~Song, C.~Meng, and S.~Ermon, ``Denoising diffusion implicit models,'' in
  \emph{Proc. Int. Conf. Learn. Represent.}, 2021.

\bibitem{nichol2021improved}
A.~Nichol and P.~Dhariwal, ``Improved denoising diffusion probabilistic
  models,'' in \emph{Proc. Int. Conf. Mach. Learn.}, 2021, pp. 8162--8171.

\bibitem{ye_3dinverse}
H.~Chung, D.~Ryu, M.~T. McCann, M.~L. Klasky, and J.~C. Ye, ``Solving 3{D}
  inverse problems using pre-trained 2{D} diffusion models,'' in \emph{Proc.
  IEEE Conf. Comput. Vis. Pattern Recognit.}, 2023.

\bibitem{ye_improving}
H.~Chung, B.~Sim, D.~Ryu, and J.~C. Ye, ``Improving diffusion models for
  inverse problems using manifold constraints,'' in \emph{Proc. Adv. Neural
  Inf. Process. Syst.}, 2021.

\bibitem{song2020score}
Y.~Song, J.~Sohl-Dickstein, D.~P. Kingma, A.~Kumar, S.~Ermon, and B.~Poole,
  ``Score-based generative modeling through stochastic differential
  equations,'' in \emph{Proc. Int. Conf. Learn. Represent.}, 2021.

\bibitem{dhariwal2021diffusion}
P.~Dhariwal and A.~Nichol, ``Diffusion models beat {GANs} on image synthesis,''
  in \emph{Proc. Adv. Neural Inf. Process. Syst.}, vol.~34, 2021, pp.
  8780--8794.

\bibitem{yang2022diffusion}
L.~Yang \emph{et~al.}, ``Diffusion models: A comprehensive survey of methods
  and applications,'' \emph{arXiv:2209.00796}, 2022.

\bibitem{gao2022cocodiff}
Q.~Gao and H.~Shan, ``{CoCoDiff}: a contextual conditional diffusion model for
  low-dose {CT} image denoising,'' in \emph{Proc. SPIE}, vol. 12242, 2022.

\bibitem{xia2022low}
W.~Xia, Q.~Lyu, and G.~Wang, ``Low-dose {CT} using denoising diffusion
  probabilistic model for 20$\times$ speedup,'' \emph{arXiv:2209.15136}, 2022.

\bibitem{bansal2022cold}
A.~Bansal \emph{et~al.}, ``Cold diffusion: {Inverting} arbitrary image
  transforms without noise,'' in \emph{Proc. Adv. Neural Inf. Process. Syst.},
  2023.

\bibitem{yen2022cold}
H.~Yen, F.~G. Germain, G.~Wichern, and J.~L. Roux, ``Cold diffusion for speech
  enhancement,'' \emph{arXiv:2211.02527}, 2022.

\bibitem{joshi2010seeing}
N.~Joshi and M.~F. Cohen, ``{Seeing Mt. Rainier}: Lucky imaging for multi-image
  denoising, sharpening, and haze removal,'' in \emph{Int. Conf. Comput.
  Photogr.}, 2010, pp. 1--8.

\bibitem{lehtinen2018noise2noise}
J.~Lehtinen \emph{et~al.}, ``{Noise2Noise}: Learning image restoration without
  clean data,'' in \emph{Proc. Int. Conf. Mach. Learn.}, 2018, pp. 2965--2974.

\bibitem{wu2019consensus}
D.~Wu, K.~Gong, K.~Kim, X.~Li, and Q.~Li, ``Consensus neural network for
  medical imaging denoising with only noisy training samples,'' in \emph{Proc.
  Int. Conf. Med. Image Comput. Comput.-Assisted Intervention}, 2019, pp.
  741--749.

\bibitem{perez2018film}
E.~Perez, F.~Strub, H.~De~Vries, V.~Dumoulin, and A.~Courville, ``{FiLM}:
  {Visual} reasoning with a general conditioning layer,'' in \emph{Proc. AAAI
  Conf. Artif. Intell.}, vol.~32, no.~1, 2018.

\bibitem{huang2017arbitrary}
X.~Huang and S.~Belongie, ``Arbitrary style transfer in real-time with adaptive
  instance normalization,'' in \emph{Proc. IEEE Int. Conf. Comput. Vis.}, 2017,
  pp. 1501--1510.

\bibitem{dumoulin2016learned}
V.~Dumoulin, J.~Shlens, and M.~Kudlur, ``A learned representation for artistic
  style,'' in \emph{Proc. Int. Conf. Learn. Represent.}, 2017.

\bibitem{li2022noise}
D.~Li, Z.~Bian, S.~Li, J.~He, D.~Zeng, and J.~Ma, ``Noise characteristics
  modeled unsupervised network for robust ct image reconstruction,'' \emph{IEEE
  Trans. Med. Imag.}, vol.~41, no.~12, pp. 3849--3861, 2022.

\bibitem{shan2019novel}
H.~Shan, U.~Kruger, and G.~Wang, ``A novel transfer learning framework for
  low-dose {CT},'' in \emph{Proc. SPIE}, vol. 11072, 2019, pp. 513--517.

\bibitem{chen2016open}
B.~Chen, S.~Leng, L.~Yu, D.~Holmes~III, J.~Fletcher, and C.~McCollough, ``An
  open library of {CT} patient projection data,'' in \emph{Proc. SPIE}, vol.
  9783, 2016, pp. 330--335.

\bibitem{zeng2015simple}
D.~Zeng \emph{et~al.}, ``A simple low-dose x-ray {CT} simulation from high-dose
  scan,'' \emph{IEEE Trans. Nucl. Sci.}, vol.~62, no.~5, pp. 2226--2233, 2015.

\bibitem{zhao2019convolutional}
T.~Zhao, M.~McNitt-Gray, and D.~Ruan, ``A convolutional neural network for
  ultra-low-dose {CT} denoising and emphysema screening,'' \emph{Med. Phys.},
  vol.~46, no.~9, pp. 3941--3950, 2019.

\bibitem{moen2021low}
T.~R. Moen \emph{et~al.}, ``Low-dose ct image and projection dataset,''
  \emph{Med. Phys.}, vol.~48, no.~2, pp. 902--911, 2021.

\bibitem{yi2018sharpness}
X.~Yi and P.~Babyn, ``Sharpness-aware low-dose {CT} denoising using conditional
  generative adversarial network,'' \emph{J Digit Imaging}, vol.~31, pp.
  655--669, 2018.

\bibitem{zhovannik2019learning}
I.~Zhovannik \emph{et~al.}, ``Learning from scanners: {B}ias reduction and
  feature correction in radiomics,'' \emph{Clin. Transl. Radiat.}, vol.~19, pp.
  33--38, 2019.

\bibitem{guo2019toward}
S.~Guo, Z.~Yan, K.~Zhang, W.~Zuo, and L.~Zhang, ``Toward convolutional blind
  denoising of real photographs,'' in \emph{Proc. IEEE Conf. Comput. Vis.
  Pattern Recognit.}, 2019, pp. 1712--1722.

\bibitem{shi2020review}
L.~Shi \emph{et~al.}, ``Review of {CT} image reconstruction open source
  toolkits,'' \emph{J. X-Ray Sci. Technol.}, vol.~28, no.~4, pp. 619--639,
  2020.

\bibitem{geng2021content}
M.~Geng \emph{et~al.}, ``Content-noise complementary learning for medical image
  denoising,'' \emph{IEEE Trans. Med. Imag.}, vol.~41, no.~2, pp. 407--419,
  2021.

\bibitem{niu2022noise}
C.~Niu \emph{et~al.}, ``Noise suppression with similarity-based self-supervised
  deep learning,'' \emph{IEEE Trans. Med. Imag.}, 2022.

\bibitem{zhang2011fsim}
L.~Zhang, L.~Zhang, X.~Mou, and D.~Zhang, ``{FSIM}: A feature similarity index
  for image quality assessment,'' \emph{IEEE Trans. Image Process.}, vol.~20,
  no.~8, pp. 2378--2386, 2011.

\bibitem{sheikh2006image}
H.~R. Sheikh and A.~C. Bovik, ``Image information and visual quality,''
  \emph{IEEE Trans. Image Process.}, vol.~15, no.~2, pp. 430--444, 2006.

\bibitem{damera2000image}
N.~Damera-Venkata, T.~D. Kite, W.~S. Geisler, B.~L. Evans, and A.~C. Bovik,
  ``Image quality assessment based on a degradation model,'' \emph{IEEE Trans.
  Image Process.}, vol.~9, no.~4, pp. 636--650, 2000.

\bibitem{mason2019comparison}
A.~Mason \emph{et~al.}, ``Comparison of objective image quality metrics to
  expert radiologists’ scoring of diagnostic quality of {MR} images,''
  \emph{IEEE Trans. Med. Imag.}, vol.~39, no.~4, pp. 1064--1072, 2019.

\bibitem{gutjahr2016human}
R.~Gutjahr \emph{et~al.}, ``Human imaging with photon-counting-based {CT} at
  clinical dose levels: Contrast-to-noise ratio and cadaver studies,''
  \emph{Invest. Radiol.}, vol.~51, no.~7, p. 421, 2016.

\bibitem{park2020enhancement}
P.~C. Park \emph{et~al.}, ``Enhancement pattern mapping technique for improving
  contrast-to-noise ratios and detectability of hepatobiliary tumors on
  multiphase computed tomography,'' \emph{Med. Phys.}, vol.~47, no.~1, pp.
  64--74, 2020.

\bibitem{van2011task}
D.~Van~de Sompel, M.~Brady, and J.~Boone, ``Task-based performance analysis of
  {FBP}, {SART} and {ML} for digital breast tomosynthesis using signal {CNR}
  and {Channelised Hotelling Observers},'' \emph{Med. Image Anal.}, vol.~15,
  no.~1, pp. 53--70, 2011.

\bibitem{wunderlich2014exact}
A.~Wunderlich, F.~Noo, B.~D. Gallas, and M.~E. Heilbrun, ``Exact confidence
  intervals for channelized {H}otelling observer performance in image quality
  studies,'' \emph{IEEE Trans. Med. Imag.}, vol.~34, no.~2, pp. 453--464, 2014.

\bibitem{yu2017correlation}
L.~Yu \emph{et~al.}, ``Correlation between a {2D} channelized {H}otelling
  observer and human observers in a low-contrast detection task with multislice
  reading in {CT},'' \emph{Med. Phys.}, vol.~44, no.~8, pp. 3990--3999, 2017.

\bibitem{kalisz2016artifacts}
K.~Kalisz, J.~Buethe, S.~S. Saboo, S.~Abbara, S.~Halliburton, and P.~Rajiah,
  ``Artifacts at cardiac {CT}: physics and solutions,'' \emph{Radiographics},
  vol.~36, no.~7, pp. 2064--2083, 2016.

\bibitem{rombach2022high}
R.~Rombach, A.~Blattmann, D.~Lorenz, P.~Esser, and B.~Ommer, ``High-resolution
  image synthesis with latent diffusion models,'' in \emph{Proc. IEEE Conf.
  Comput. Vis. Pattern Recognit.}, 2022, pp. 10\,684--10\,695.

\bibitem{song2023consistency}
Y.~Song, P.~Dhariwal, M.~Chen, and I.~Sutskever, ``Consistency models,'' in
  \emph{Proc. Int. Conf. Mach. Learn.}, 2023.

\end{thebibliography}
\end{document}